\documentclass[lettersize,journal]{IEEEtran}
\usepackage{graphicx}
\usepackage{cite}
\usepackage{amsmath,amsfonts}
\usepackage{multirow}
\usepackage{makecell}
\usepackage{threeparttable} 
\usepackage{booktabs}%
\usepackage[ruled]{algorithm}
\usepackage{algorithmic}
\usepackage[caption=false,font=normalsize,labelfont=sf,textfont=sf]{subfig}
\usepackage{stfloats}
\usepackage{epstopdf}
\usepackage{color}

\begin{document}

\title{SOM-Net: Unrolling the Subspace-based Optimization for Solving Full-wave Inverse Scattering Problems}

\author{Yu Liu \IEEEmembership{Member, IEEE}, Hao Zhao, Rencheng Song \IEEEmembership{Member, IEEE}, Xudong Chen \IEEEmembership{Fellow, IEEE}, Chang Li \IEEEmembership{Member, IEEE} and Xun Chen, \IEEEmembership{Senior Member, IEEE}
        % <-this % stops a space
\thanks{This work was supported  in part by the National Natural Science Foundation of China (Grants 62176081, 62271186, 61922075, and 41901350), the Provincial Natural Science Foundation of Anhui (Grant 2108085MF201, 2008085QF285), the Anhui Key Project of Research and Development Plan (Grants  202104d07020005),
and  the Fundamental Research Funds for the Central Universities (Grants PA2021KCPY0051, JZ2020HGPA0111 and JZ2021HGTB0078).  (Corresponding
author: Rencheng Song).}%
\thanks{Yu Liu, Hao Zhao, Rencheng Song and Chang Li  are with the Department of Biomedical Engineering, and also with the Anhui Province Key Laboratory of Measuring Theory and Precision Instrument, Hefei University of Technology, Hefei 230009, China (e-mail:	
yuliu@hfut.edu.cn; haozhao1@mail.hfut.edu.cn; rcsong@hfut.edu.cn; changli@hfut.edu.cn).}
\thanks{Xudong Chen is with the  Department of Electrical and Computer Engineering, National University of Singapore, 117583, Singapore
(e-mail: elechenx@nus.edu.sg).}
\thanks{Xun Chen is with the  Department of Electronic Engineering and Information Science,
University of Science and Technology of China, Hefei, Anhui, 230027,
China  (e-mail: xunchen@ustc.edu.cn). }}

\markboth{Journal of \LaTeX\ Class Files,~Vol.~14, No.~8, August~2021}%
{Shell \MakeLowercase{\textit{et al.}}: A Sample Article Using IEEEtran.cls for IEEE Journals}


\maketitle

\begin{abstract}
In this paper, an unrolling algorithm of the iterative subspace-based optimization method (SOM) is proposed for solving full-wave inverse scattering problems (ISPs). The unrolling network, named SOM-Net,  inherently embeds the Lippmann-Schwinger physical model into the design of network structures. The SOM-Net takes the deterministic induced current and the raw permittivity image obtained from back-propagation (BP) as the input. It then updates the induced current and the permittivity successively in sub-network blocks of the SOM-Net by imitating iterations of the SOM. The final output of the SOM-Net is the full predicted induced current, from which the scattered field and the permittivity image can also be deduced analytically. The parameters of the SOM-Net are optimized in a supervised manner with the total loss to simultaneously ensure the consistency of the induced current, the scattered field, and the permittivity in the governing equations. Numerical tests on both synthetic and experimental data verify the superior performance of the proposed SOM-Net over typical ones. The results on challenging examples like scatterers with tough profiles or high permittivity demonstrate the good generalization ability of the SOM-Net. With the use of deep unrolling technology, this work  builds a bridge between traditional iterative methods and deep learning methods for solving ISPs.

\end{abstract}


\section{Introduction}
\label{sec:introduction}
\IEEEPARstart{F}ull-wave electromagnetic inverse scattering problems (ISPs) aim  at determining the shape, position, and constitutive parameters of unknown scatterers from the measured scattering field. It is well known that the full-wave ISP is usually highly nonlinear and ill-posed \cite{chen2018computational}, which needs to be solved by nonlinear iterative methods based on governing physical models. Typical nonlinear iterative ISP methods include the distorted Born iterative method (DBIM) \cite{chew1990reconstruction}, the contrast source inversion (CSI) method \cite{van1997contrast}, the contrast source extended Born (CS-EB) method \cite{crocco2005testing}, and the subspace-based optimization method (SOM) \cite{chen2009subspace,xu2017hybrid}, etc. Those methods have achieved great success in reconstructing scatterers in various ISP applications. However, one main bottleneck of nonlinear iterative methods is the lack of real-time reconstruction performance due to the expensive computational cost. In addition, the reconstruction quality is also highly dependent on the complicated regularization.

In recent years, researchers have successfully introduced deep learning (DL) technology to solve ISPs \cite{massa2019dnns,chen2020review,li2021machine}. Many studies demonstrate that DL-based methods outperform conventional iterative ones for both reconstruction speed and accuracy. Nevertheless, a major concern of DL-based methods is the generalization ability of the learned models. Namely, the performance of DL-based methods may degrade significantly, when the testing target is quite different from those training ones on the aspects such as shape, size, and parameter ranges. This is because DL models are typically data-driven, and the constraints of the governing physical laws are only partially satisfied compared to existing iterative full-wave methods.

Many efforts have been made to improve the model generalization ability of DL-based inversion methods, which are often referred to as physics-inspired models. In general, existing physics-inspired methods can be divided into two types according to the different ways of model learning. The first type of methods accelerate  the conventional iterative methods using the DL technique, which only adopt  neural networks to replace some expensive components in the traditional objective function method.

For example, Guo et al. \cite{guo2019supervised} introduced a supervised descent method (SDM) to learn the average descent directions from data. In \cite{guo2020pixel}, Guo et al. applied the SDM to the pixel-based and model-based inversion of microwave data for dielectric targets. This type of methods have a great advantage on generalization ability, but the real-time performance is still a challenge.

The second type of physics-inspired methods are composed of two steps, i.e., a rough permittivity image reconstruction by a fast imaging method, followed by a  permittivity image enhancement to get a high-resolution target \cite{li2018deepnis,sun2018efficient}.
Among these methods, the back-propagation (BP) method is commonly used to  reconstruct the rough permittivity image from the scattered field. Then, the resolution of rough input image is enhanced in an image-to-image translation way using various networks, such as the U-Net \cite{ronneberger2015u}, GANs \cite{goodfellow2014generative,ye2020inhomogeneous}, cGANs \cite{isola2017image}, etc. The performance of this kind of methods is dominated by both the quality of input images and the way of conducting image resolution enhancement.
Wei et al. \cite{wei2018deep} proposed the dominant current scheme (DCS) instead of BP to generate a better rough input for the U-Net. Zhang et al. \cite{zhang2020learning} improved the input by combining the rough images from both qualitative and quantitative methods.   Yao et al. \cite{yao2019two} proposed to obtain the initial input image using a shallow convolutional neural network (CNN) and then enhance the resolution using  U-Net.
In addition to these efforts to improve the quality of input images, Huang et al. \cite{huang2020deep} adopted the  structural similarity index measure (SSIM) loss and the mean squared error (MSE) loss to jointly constrained image features and improve the reconstruction quality. Song et al. \cite{song2021electromagnetic} defined a novel perceptual adversarial network to improve the quality of image-to-image translation. Wei et al. \cite{wei2019physics}  introduced an induced current learning method (ICLM) to solve the translation in the contrast source domain instead of the contrast domain.
Although these methods achieve better reconstructions, the image-to-image translation in ISP is still lacking of physical insight,   thereby restricting the generalization ability of this kind of DL-based ISP methods. Specifically, the image-to-image translation in existing ISP methods only restricts the match of a single physical variable, which is usually the permittivity. The ignoring of the consistency of other physical parameters may also degenerate the model generalization ability.

Recently, the unrolling technique \cite{monga2021algorithm} has attracted extensive attention in image processing and computational imaging research, which builds a bridge between  conventional iterative methods and DL-based methods. The deep unrolling technique constructs neural networks with a few sub-network modules to imitate iterations of conventional methods. The unknown target parameters are successively updated  in each sub-network module like doing iterations. It effectively embeds physical knowledge into deep neural networks, which makes the DL model more interpretable and efficient. The unrolling technique has been verified to be efficient in solving various tasks like medical imaging \cite{jin2017deep,adler2017solving,yang2018admm}, and image restoration \cite{li2020efficient,schuler2015learning}, etc.
It has also been introduced to solve nonlinear ISPs very recently. Liu et al. \cite{liu2021physical} introduced the PM-Net to simulate the ADMM iterative process, which defines four unconstrained sub-problems to achieve  the alternating optimization of contrast source and contrast. Guo et al. \cite{guo2021physics} proposed an ISP SOLVER to simulate the iterative gradient-based inversion method, where a pre-trained forward  model is used repeatedly to calculate the data fitness. The performance of the ISP SOLVER is strongly affected by that of the pre-trained FWD SOLVER model. These methods have been verified to obtain good generalization capability compared to direct data-to-model networks. However, only partial physical variables are explicitly enforced to satisfy the consistency in these methods. The embedding of the governing physical law in learned networks can still be further improved.

In this work, an unrolling of the subspace-based optimization, named SOM-Net, is proposed to solve  nonlinear ISPs. The input of the SOM-Net is taken as the deterministic induced current obtained by the singular value decomposition (SVD) and the raw permittivity image obtained from BP, while the output of SOM-Net is the predicted full induced current. The induced current and the permittivity image are updated alternatively by passing through the sub-network blocks in SOM-Net, which is like performing successive iterations in SOM. Particularly, the update of the induced current in the SOM-Net is realized by the mapping of each sub-network, while the permittivity is still updated analytically. Finally, the scattered field and  the permittivity are also deduced analytically from the output induced current. The whole SOM-Net model is optimized with residuals defined simultaneously on the induced current, the scattered field, and the permittivity image in a supervised manner. The match of all these physical parameters which are governed by the Lippmann-Schwinger equation ensures  the generalization ability of the SOM-Net model. Numerical tests on both synthetic and experimental data verify the superior performance of the proposed SOM-Net over existing ones.

In summary, the contributions of this work are listed as follows:
\begin{itemize}
\item The proposed SOM-Net inherently embeds physical knowledge by unrolling SOM, which introduces an effective way to design physics-inspired deep learning networks from the conventional iterative methods. Specifically, the update of the induced current and the permittivity image within each sub-network of the SOM-Net effectively reduces the nonlinearity of the mapping. Besides, the  permittivity is always updated analytically in SOM-Net which greatly simplifies the network structure.  Therefore, the SOM-Net is physically interpretable and  efficient.
\item  The full constraints of physical variables, including the induced current, the scattered field, and the permittivity image, make  the output of the SOM-Net physically consistent with the governing law. Guided by those constraints, the prediction of the SOM-Net is more reliable and consistent, in terms of not only the reconstructed images but also the induced current and the scattered field.
\item The testing results on various challenging examples  demonstrate the effectiveness and good generalization ability of the SOM-Net. In particular, the comparison results on scatterers with tough profiles or high permittivity verify the advantages of the proposed method over other comparable ones in terms of both accuracy and generalization ability.
\end{itemize}

This paper is organized as follows. In Section II, we introduce the formulations of the proposed algorithm.  Synthetic and experimental data are then tested in Section III for the performance verification. In  Section IV, we discuss the benefit of SOM-net and its applications  in other scenarios.  Finally, we conclude our work   in Section V.

\begin{figure}[ht]
\centering
\includegraphics[width=0.28\textwidth]{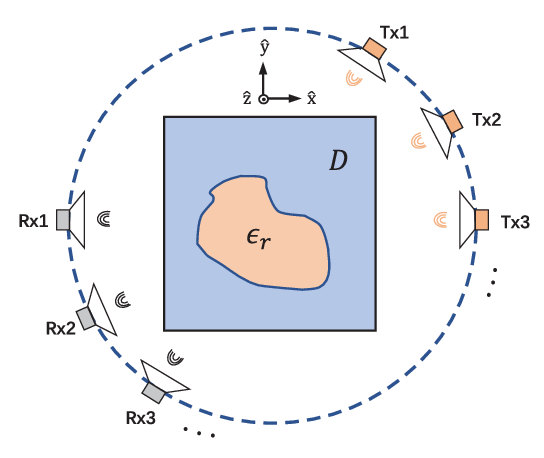}\\
\caption{{The configuration of a 2-D ISP under TM illuminations.}}
\label{TxRx}
\end{figure}

\section{Method}
For clarity, the notations are declared first. We use $\bar{\bar{X}} $ and $\bar{X} $ to denote the matrix and vector of the discretized operator or parameter $X$, respectively. The superscripts $H$ and $*$ respectively indicate the conjugate transpose and the complex conjugate of a matrix or vector.
\subsection{Formulation of the Problem}
The configuration of ISPs is shown in Fig. \ref{TxRx}. For convenience, a 2-D transverse magnetic (TM) \cite{li2017probabilistic} wave is considered, where the longitude direction is along $\hat{z}$. The background is free-space with permittivity $\epsilon_{0}$ and permeability $\mu _{0}$. The unknown scatterers are { supposed to be lossless by default and} located in the domain of interest (DOI) $D $. The $N_{i} $ transmitters located at $r_{l} $ with $ l= 1,2,...,N_{i} $ sequentially illuminate the unknown scatterers, and the $N_{r} $ receivers located at $r_{q} $ with $q= 1,2,...,N_{r} $ simultaneously measure the scattered field.

\subsubsection{Forward Scattering}
The forward scattering aims to calculate scattering field with known permittivity of scatterers.
The method of moments (MoM) \cite{peterson1998computational} is taken to calculate the scattered field. The domain $D$ is discretized into $M$ subunits, and the centers of subunits are located at $r_{m}$ with $m=1,2,...M$. Accordingly, taking into account of all subunits, the state equation is discretized as
\begin{equation}\label{Eq3}
 \bar{{E}}^{t} =\bar{{E}}^{i} +\bar{\bar{{G}}}_{D} \cdot \bar{J},
\end{equation}
where $\bar{{E}}^{i} $ and $\bar{{E}}^{t}$ represent the incident and total electric field in domain $D$, respectively. The matrix $\bar{\bar{{G}}}_{D}$ with $M\times M$ dimensions maps the induced current to the scattered field in domain $D$. In the forward problem, \eqref{Eq3} describes the wave-scattering interaction in domain $D$.

The induced current $\bar{J}$ is defined as
$\bar{J}=\bar{\bar{{\chi}}}\cdot \bar{{E}}^{t}$. It is well known that the contrast $\bar{\bar{\chi}}$ and the relative permittivity $\bar{\bar{\epsilon }}_{r}$ have a relationship of $\bar{\bar{{\chi}}}=-i\left ( k_{0}/\eta _{0} \right ) \left ( \bar{\bar{\epsilon }}_{r}  -1 \right ) $, where $i=\sqrt{-1} $, $k_{0} $ and $\eta _{0} $ respectively represent the wavenumber and the impedance of homogeneous medium background.

The discretized data equation is expressed as
\begin{equation} \label{Eq4}
\bar{{E}}^{s} =\bar{\bar{{G}}}_{S} \cdot \bar{J},
\end{equation}
where $\bar{{E}}^{s}$ is the scattered field measured by all receivers. The matrix $\bar{\bar{{G}}}_{S}$ with $N_{r}\times M$ dimensions maps the induced current from the domain $D$ to the space of scattered field in domain $S$. In the forward problem, \eqref{Eq4} describes the scattered field as a reradiation of the induced current.

\subsubsection{SOM Algorithm}
The purpose of the inverse problem is to retrieve the relative permittivity $\bar{\bar{\epsilon }}_{r}$ of the scatterer from the measured scattering field $\bar{{E}}^{s}$.
For solving nonlinear ISPs, the induced current cannot be obtained directly from \eqref{Eq4}. Suppose the singular value decomposition (SVD) of $\bar{\bar{{G}}}_{S}$ is $\bar{\bar{{G}}}_{S}= {\textstyle \sum_{n}} \bar{u}_{n}\sigma_{n}\bar{\nu}_{n}^{\small H}$ with $\sigma _{1}\ge\sigma _{2}\ge,..., \sigma _{M}=0 $, where $\bar{\mu}_{n}$, $\bar{\nu}_{n}$ and $\sigma_{n}$ are the orthogonal left, right singular vectors, and singular values of $\bar{\bar{{G}}}_{S}$, respectively. The singular values are sorted in the descending order.

In SOM, the full induced current $\bar{J}$ is decoupled into the deterministic induced current $\bar{J}^{+}$ and the ambiguous induced current $\bar{J}^{-}$ as
\begin{equation} \label{Jfull}
\bar{J} = \bar{J}^{+}+\bar{J}^{-} = \bar{J}^{+}+\bar{\bar{V}}^{-}\cdot \bar{\alpha}^{-},
\end{equation}
where $\bar{\bar{V}}^{-}$ is composed of the last $M-L$ columns of right singular vectors, $\bar{\alpha}^{-}$ is the unknown coefficient vector with the length of $M-L$, and $\bar{{J}}^{+}$ satisfies
 \begin{equation} \label{Eq7}
\bar{{J}}^{+}=\sum_{j=1}^{L}\frac{\bar{\mu}_{j}^{\scriptsize H}\cdot \bar{E}^{s} }{\sigma_{j}} \bar{\nu}_{j}.
\end{equation}

The initializations of SOM are taken as the deterministic induced current $\bar{{J}}^{+}$ and the contrast $\bar{\bar{\chi}}^{bp}$ obtained by back-propagation in \eqref{bp},
\begin{equation} \label{bp}
\bar{\bar{\chi}}^{bp}(r)  =\frac{ {\textstyle \sum_{l=1}^{N_{i} }\bar{J} _{l}(r)\cdot \left [ \bar{E} _{l}^{t}(r)  \right ]^{*}   } }{ {\textstyle \sum_{l=1}^{N_{i} }\left | \bar{E} _{l}^{t}(r) \right |^{2}  } }.
\end{equation}

In SOM, the unknown parameters, $\bar{\alpha}^{-} $ and $\bar{\bar{\chi}} $ are updated alternatively. For the $m $th subunits, the induced current $(\bar{J}_{l})_{m}$ is updated by \eqref{Jfull}
, the total field $(\bar{E}^{t}_{l})_{m}$ is updated by \eqref{Eq3}
and the contrast $(\bar{\bar{\chi}})_{m} $ is straightforwardly updated by \eqref{chai} \cite{chen2018computational},
\begin{equation} \label{chai}
(\bar{\bar{\chi}})_{m}=\left [ \sum_{l=1}^{N_{i} }\frac{(\bar{E}_{l}^{t}  )_{m}^{*} }{\left \| \bar{{J}}_{l}^{+} \right \| } \cdot \frac{(\bar{J}_{l}  )_{m}}{\left \| \bar{{J}}_{l}^{+} \right \| }  \right ] / \left [ \sum_{l=1}^{N_{i} }\left | \frac{(\bar{E}_{l}^{t}  )_{m} }{\left \| \bar{J}_{l}^{+} \right \| } \right | ^{2} \right ],
\end{equation}
where ${N_{i} }$ is the total number of  transmitters, which is set as 16.

\subsection{The SOM-Net Algorithm}

\begin{figure*}[ht]
\centering
 \setlength{\abovecaptionskip}{-2pt}
\includegraphics[width=0.9\textwidth]{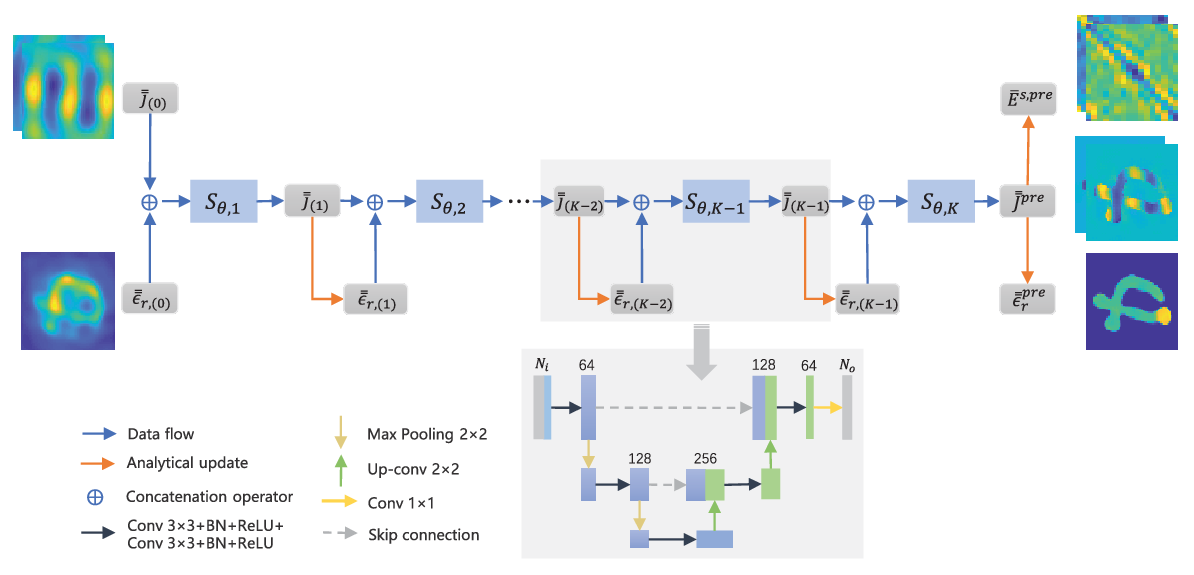}\\
\caption{{The structure of the proposed SOM-Net.} The input of the SOM-Net is a combined variable with 3 input channels, which contains the deterministic induced current and the raw permittivity image obtained from BP. The output of the SOM-Net is the predicted full induced current $\bar{\bar{J}}^{pre} $ with 2 output channels, which is further used to calculate the predicted scattered field $\bar{E}^{s,pre} $ and the predicted permittivity image $\bar{\bar{\epsilon}}_{r}^{pre}$ through \eqref{Eq4} and \eqref{chai}, respectively. The superscript $pre$ indicates the variable is obtained from prediction. The symbol $\oplus$ denotes the concatenation operation for the input channels. The orange arrows denote the variables are updated analytically.}
\label{flowchart}
\end{figure*}

\subsubsection{Structure of SOM-Net}
The flowchart of the proposed SOM-Net is shown in Fig. \ref{flowchart}. The SOM-Net is designed by unrolling the iterations in SOM, which not only incorporates physical knowledge into the model, but also alleviates the nonlinear coupling of the induced current and the permittivity.

As seen, there are $K$ sub-networks in SOM-Net to imitate the SOM iterations. Suppose $\bar{\bar{\epsilon}} _{r}^{bp}$ denotes the permittivity image reconstructed by the BP method with all scattered field. The $\bar{\bar{J}}^{+}$ represents the  deterministic induced current matrix reformulated from the vector $\bar{{J}}^{+}$. For the $l$th incidence, the deterministic induced current $\bar{\bar{J}}^{+}_{l}$ is concatenated with the permittivity image $\bar{\bar{\epsilon}} _{r}^{bp}$ as a single input sample of the SOM-Net.
Accordingly, the input of the SOM-Net model is denoted as $\bar{\bar{J}}^{+}\oplus \bar{\bar{\epsilon}} _{r}^{bp} $ with the number of input channels as 3, in which the first two channels are occupied by the real and imaginary part of the induced current and the third channel is occupied by the permittivity image, and the symbol $\oplus$ indicates the concatenation operation of two variables on the real and imaginary channels.  For the same reconstruction, all the $N_{i}$ resulting samples from the total $N_{i}$ incidences will be simultaneously employed to train the SOM-Net. Namely, the batch size is set to $N_{i}$.

Similar to SOM, the SOM-Net updates the induced current $\bar{\bar{J}}_{(k)}$ and the permittivity variable $\bar{\bar{\epsilon}} _{r,(k)}$  alternatively by passing variables across the sub-network  $S_{\theta,k}, k=1,2,\ldots, K$, where $\theta $ indicates model parameters.
Finally, the predicted induced current $\bar{\bar{J}}^{pre}$ with the channel number as 2 at all incidence is directly obtained by network reconstruction. Then, the predicted scattered field $\bar{E}^{s,pre} $ is calculated analytically by \eqref{Eq4} and the permittivity image $\bar{\bar{\epsilon}}_{r}^{pre} $ can also be reconstructed directly from the contrast $\bar{\bar{\chi}}^{pre}$ by \eqref{chai}.
In SOM-Net, the analytical update of physical variables is indicated by  orange arrows in Fig. \ref{flowchart}. It significantly simplifies the network complexity and embeds physical knowledge into the model.

Overall, the data flow in SOM-Net is summarized in Algorithm \ref{algorithm}, where  $F$ represents the analytical update of permittivity from the contrast $\bar{\bar{\chi}}$ by \eqref{chai}. Specifically, four sub-networks (i.e., $K=4$) are experimentally taken to design SOM-Net in this paper. It is worth noting that $K$ can be adjusted according to  the complexity of the target ISP.

\begin{algorithm}[htbp]
\caption{Data Flow in SOM-Net}
\label{algorithm}
\begin{algorithmic}[1]
\STATE Calculate $\bar{{E}}^{i}$,$\bar{\bar{{G}}}_{D}$,$\bar{\bar{{G}}}_{S}$;\\
\STATE Obtain network input $\bar{\bar{J}}_{(0)} \gets \bar{\bar{J}}^{+}$, $\bar{\bar{\epsilon}} _{r,(0)} \gets \bar{\bar{\epsilon}} _{r}^{bp}$; \\
\FOR{each $k\in [2,K]$}
\STATE $ \bar{\bar{J}}_{(k-1)} = S_{\theta,k-1}\left ( \bar{\bar{J}}_{(k-2)}\oplus\bar{\bar{\epsilon}} _{r,(k-2)} \right ) $; \\
\STATE $ \bar{\bar{\epsilon}} _{r,(k-1)} = F\left ( \bar{\bar{J}}_{(k-1)}\right ) $;\\
\ENDFOR
\STATE $ \bar{\bar{J}}^{pre}=\bar{\bar{J}}_{(K)}$;\\
\STATE $ \bar{E}^{s,pre} = \bar{\bar{{G}}}_{S} \cdot \bar{\bar{J}}^{pre}$;\\
\STATE $ \bar{\bar{\epsilon}} _{r}^{pre}= \bar{\bar{\epsilon}} _{r,(K)}$;\\
\end{algorithmic}
\end{algorithm}

\subsubsection{Loss Functions}
In this paper,  a comprehensive loss function is defined to train the SOM-Net $S_{\theta }$.
The full loss function $L_{S}(\theta )$ of SOM-Net is defined as
\begin{equation} \label{fullloss}
L_{S}(\theta ) = L_{J}(\theta )+L_{E}(\theta )+\lambda _{1} L_{SSIM}(\theta )+\lambda _{2}L_{MSE}(\theta ),
\end{equation}
where $L_{J}(\theta )$ denotes the induced current loss, $L_{E}(\theta )$ is the scattered field loss, $L_{SSIM}(\theta )$ indicates the structural similarity loss of two permittivity images, and $L_{MSE}(\theta )$ is the pixel-wise MSE loss. Here,  $\lambda _{i} (i=1,2)$ are hyperparameters.

Specifically, each term of $L_{S}(\theta )$ is defined as

\begin{equation} \label{lossJ}
L_{J}=\frac{1}{N_{i}} \sum_{l=1}^{N_{i} }\left |\bar{\bar{J}}^{pre}_{l}-\bar{\bar{J}}^{\mathrm{MoM}}_{l} \right |^{2},
\end{equation}
\begin{equation} \label{lossE}
L_{E}=\frac{1}{N_{r}} \sum_{q=1}^{N_{r} }\left |\bar{E}^{s,pre}_{q}-\bar{E}^{s}_{q}\right |,
\end{equation}
\begin{equation} \label{lossSSIM}
L_{SSIM}=1-\mathrm{SSIM}(\bar{\bar{\epsilon}} _{r}^{pre},\bar{\bar{\epsilon }}_{r}^{t} ),
\end{equation}
\begin{equation} \label{lossMSE}
L_{MSE}=\left |\bar{\bar{\epsilon}} _{r}^{pre}-\bar{\bar{\epsilon }}_{r}^{t} \right | ^{2},
\end{equation}
where $\bar{\bar{J}}^{pre} $, $\bar{E}^{s,pre} $ and $\bar{\bar{\epsilon}} _{r}^{pre} $ are the predicted induced current, the predicted scattered field and the permittivity image, respectively, while $\bar{\bar{J}}^{\mathrm{MoM}}$, $\bar{E}^{s}$ and $\bar{\bar{\epsilon}} _{r}^{t}$ are the corresponding reference variables, respectively.

\begin{figure*}[!t]
  \centering
  \setlength{\abovecaptionskip}{-1pt}
  \includegraphics[width=0.7\textwidth]{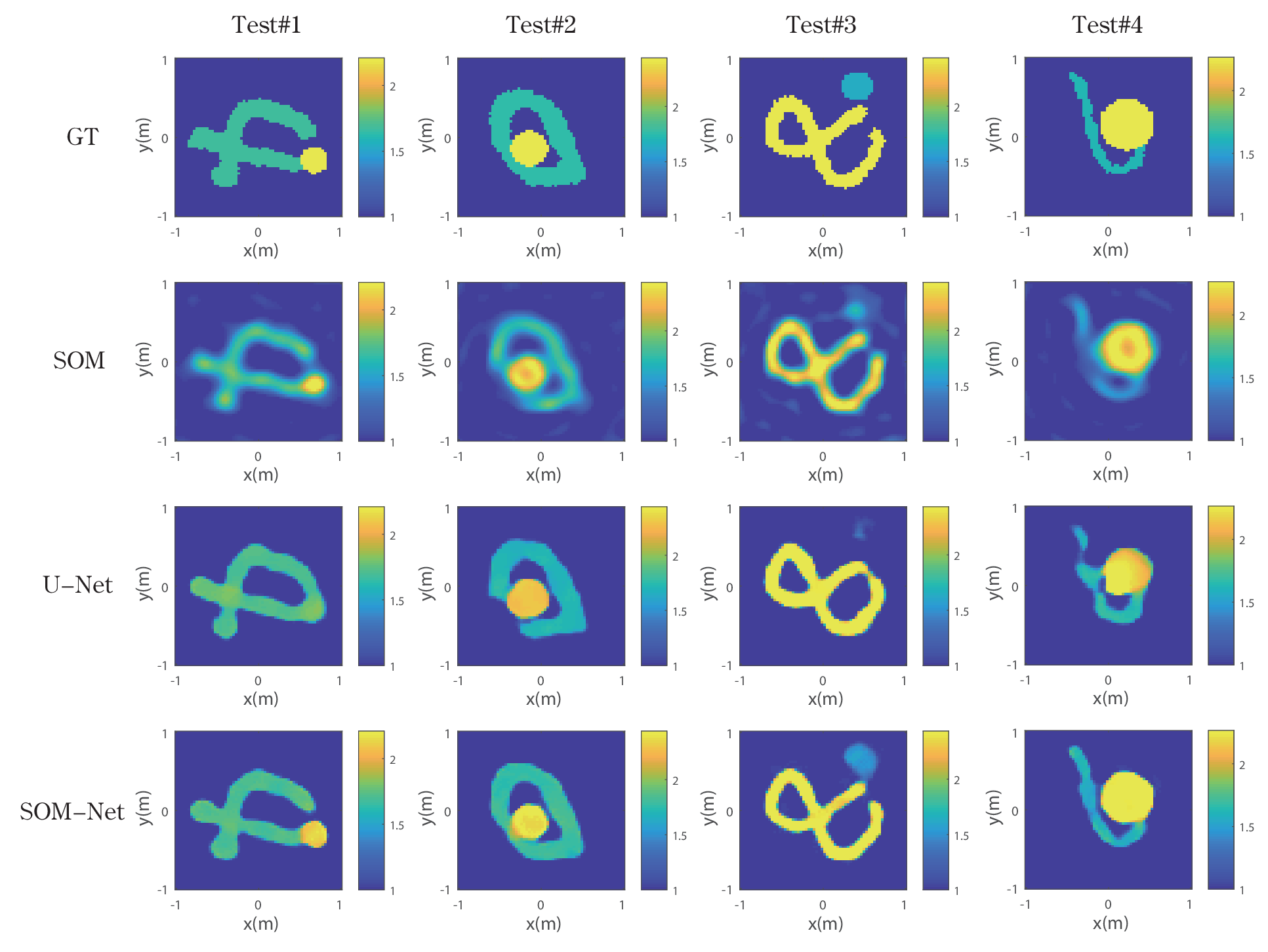}\\
  \caption{Reconstruction results of Test\#1 to Test\#4 from the MNIST data set with 10\% white Gaussian noise.}
  \label{digits4}
\end{figure*}

% 1500 MNIST TESTS
\begin{table*}[ht]
\centering
\setlength{\abovecaptionskip}{-1pt}
\caption{Error metrics of reconstruction results for 1500 testing samples from MNIST data set  by SOM, U-Net, and SOM-Net in Fig.\ref{digits4}.}
\begin{tabular}{|c|cc|cc|cc|cc|cc|}
\hline
\multirow{2}{*}{Method} & \multicolumn{2}{c|}{Test\#1}                       & \multicolumn{2}{c|}{Test\#2}                       & \multicolumn{2}{c|}{Test\#3}                       & \multicolumn{2}{c|}{Test\#4}                       & \multicolumn{2}{c|}{1500 MNIST}                    \\ \cline{2-11}
                        & \multicolumn{1}{c|}{SSIM}          & RMSE          & \multicolumn{1}{c|}{SSIM}          & RMSE          & \multicolumn{1}{c|}{SSIM}          & RMSE          & \multicolumn{1}{c|}{SSIM}          & RMSE          & \multicolumn{1}{c|}{SSIM}          & RMSE          \\ \hline
SOM                     & \multicolumn{1}{c|}{0.84}          & 0.09          & \multicolumn{1}{c|}{0.84}          & 0.11          & \multicolumn{1}{c|}{0.77}          & 0.18          & \multicolumn{1}{c|}{0.83}          & 0.12          & \multicolumn{1}{c|}{0.80}          & 0.13          \\ \hline
U-Net                   & \multicolumn{1}{c|}{0.89}          & 0.09          & \multicolumn{1}{c|}{0.88}          & 0.11          & \multicolumn{1}{c|}{0.84}          & 0.20          & \multicolumn{1}{c|}{0.87}          & 0.10          & \multicolumn{1}{c|}{0.88}          & 0.13          \\ \hline
SOM-Net                  & \multicolumn{1}{c|}{\textbf{0.92}} & \textbf{0.07} & \multicolumn{1}{c|}{\textbf{0.94}} & \textbf{0.07} & \multicolumn{1}{c|}{\textbf{0.88}} & \textbf{0.14} & \multicolumn{1}{c|}{\textbf{0.95}} & \textbf{0.07} & \multicolumn{1}{c|}{\textbf{0.91}} & \textbf{0.10} \\ \hline
\end{tabular}
\label{tab1}
\end{table*}

It can be seen that the total loss $L_{S}(\theta )$ enforces the consistency of the induced current, the scattered fiel, and the permittivity.  The use of SSIM and MSE constraints on the permittivity image further ensures the structural and pixel-wise match with the reference one.  Hence, from the physical point of view, the total loss $L_{S}(\theta )$ guarantees the consistency of all physical variables in the Lippmann-Schwinger  governing equation, thereby improving the prediction accuracy of the SOM-Net model. The proper design of network and loss functions enables the model to learn physical laws from the data, thereby making the model have good generalization ability.

\begin{figure*}[htbp]
  \centering
  \setlength{\abovecaptionskip}{-2pt}
  \includegraphics[width=0.8\textwidth]{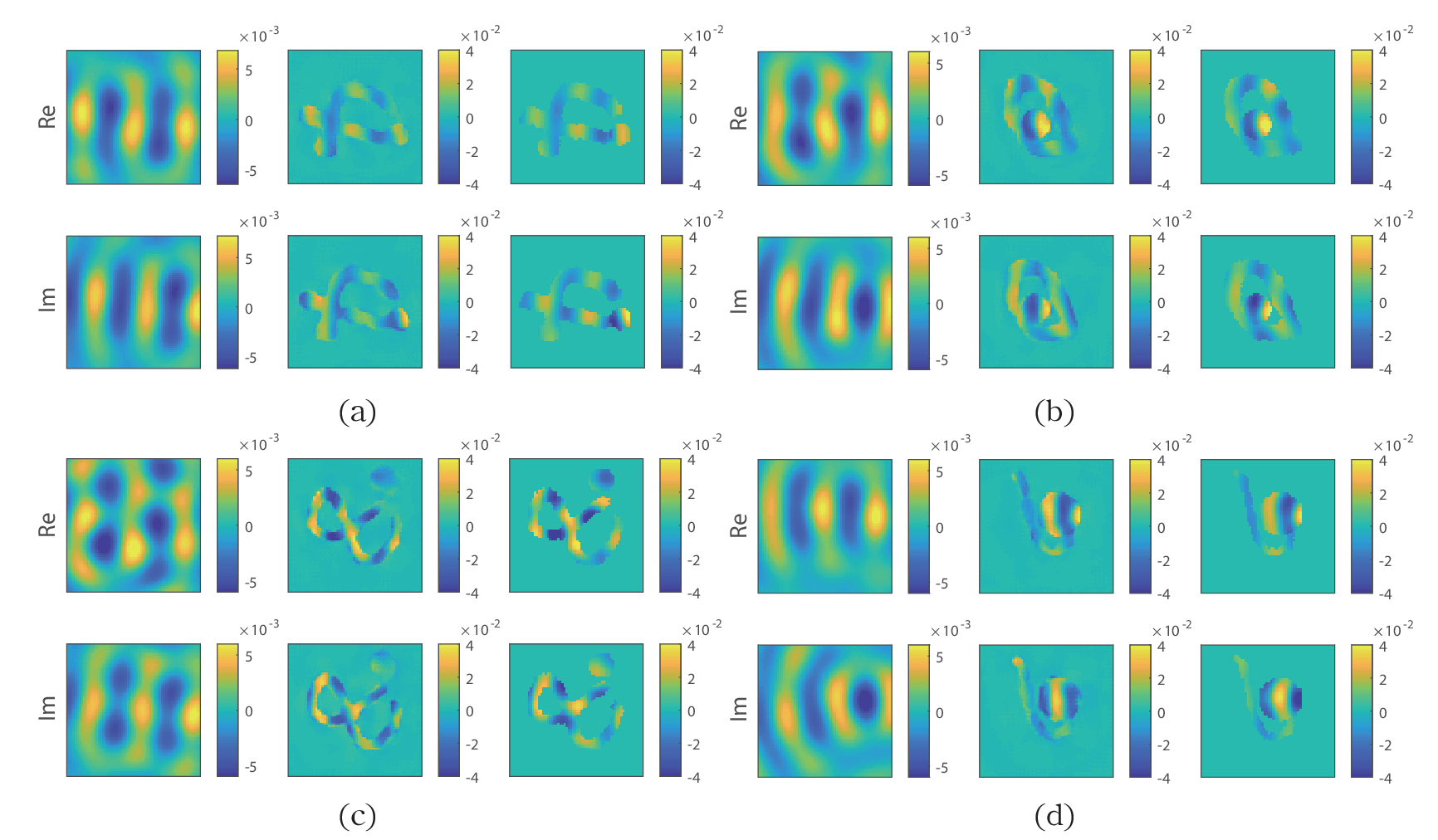}\\
  \caption{The fitting of  induced currents by the first transmitter for Test\#1 to Test\#4: (a) Test\#1, (b) Test\#2, (c) Test\#3, (d) Test\#4. For each test,  the left, middle, and right columns are the deterministic induced current, the output induced current of SOM-Net, and the target induced current, respectively. And the first row is the real part (Re), and the second row is the imaginary part (Im).}
  \label{digits4J}
\end{figure*}

\begin{figure*}[htbp]
  \centering
  \setlength{\abovecaptionskip}{-2pt}
  \includegraphics[width=1\textwidth]{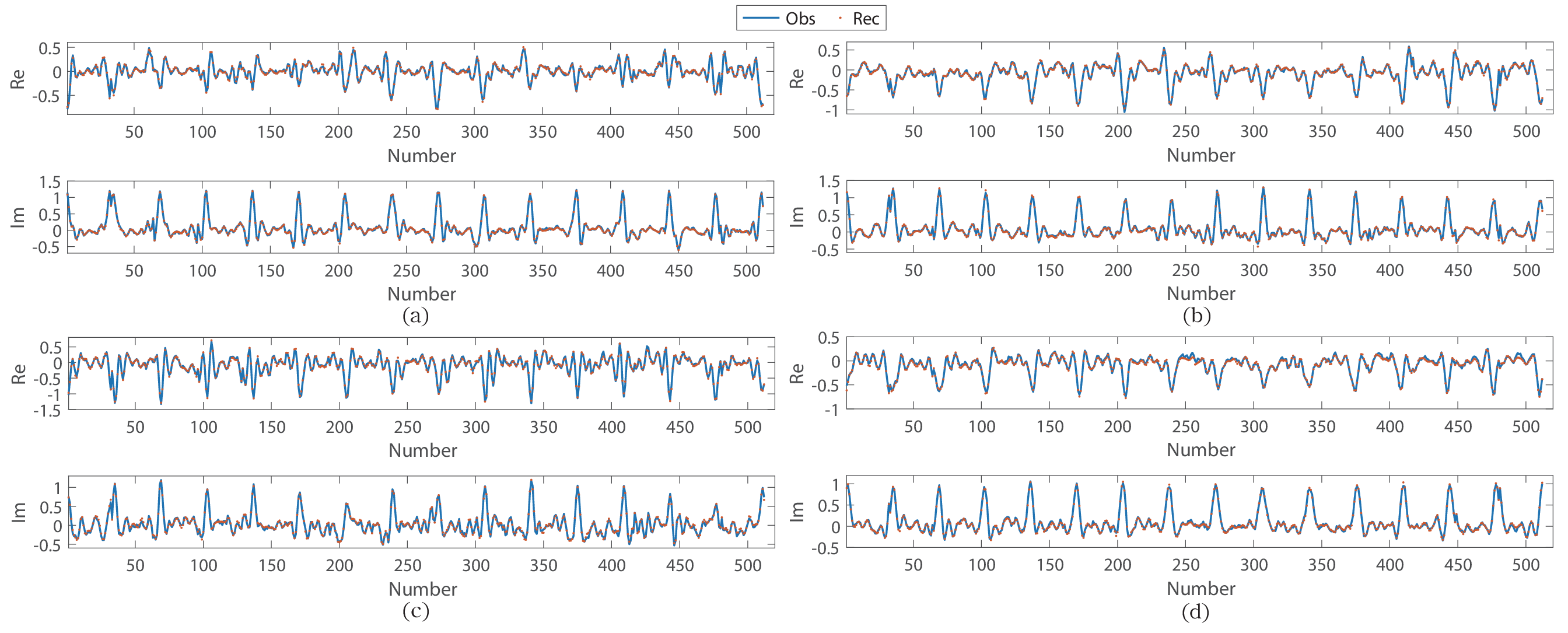}\\
  \caption{The fitting of all scattered fields for Test\#1 to Test\#4: (a) Test\#1, (b) Test\#2, (c) Test\#3, (d) Test\#4. For each test, the predicted scattered field by SOM-Net is represented by the orange dots, and the reference one is indicated by the blue line.}
  \label{digits4ES}
\end{figure*}

\subsection{Computational Complexity}
The computational complexity of the SOM-Net is mainly divided into three parts, preparing input data, feed-forward calculation of SOM-Net, and the estimation of the scattered field and the permittivity.

To prepare input data, it needs to  obtain the  raw permittivity image $\bar{\bar{\epsilon }}_{r}^{bp}$  by  BP and get the deterministic induced current $\bar{{J}}^{+}$. The computational complexity of BP is $ O\left ( N_{i}M\log{M} \right )$ \cite{wei2018deep} if the Fast Fourier transform (FFT) is applied in the matrix-vector multiplication, and $N_{i}$ is the number of incidences.
The computational cost of obtaining $\bar{{J}}^{+}$ is $ O\left ( N_{r}^{2} M \right ) $ \cite{stewart1998matrix}, which lies in the thin SVD decomposition of $\bar{\bar{{G}}}_{S} $.

For the feed-forward of SOM-Net, the computational cost includes several basic operations like convolutions, activation function, and max pooling, where the complexity is dominated by convolutions. Especially, assume that when performing convolution operation, the number of input feature maps is $Q_{i}$ and the number of output feature maps is $Q_{o}$. Thus, the complexity is calculated as $O\left ( Q_{i}Q_{o} M_{1}M_{2}K_{f}^{2} \right ) $  \cite{wei2018deep}, where the feature map size and the convolution kernel size are $M_{1} \times M_{2}$ and $K_{f} \times K_{f} $ ($K_{f} = 3$ in this paper), respectively.

To calculate the scattered field, the computational cost of the matrix-vector multiplication in \eqref{Eq4} is $O\left ( N_{r}M\right )$. To calculate the permittivity, the computational cost of the vector-vector multiplication in \eqref{chai} is $O\left (M\right )$.

\begin{figure*}[!t]
  \centering
  \setlength{\abovecaptionskip}{-1pt}
  \includegraphics[width=1\textwidth]{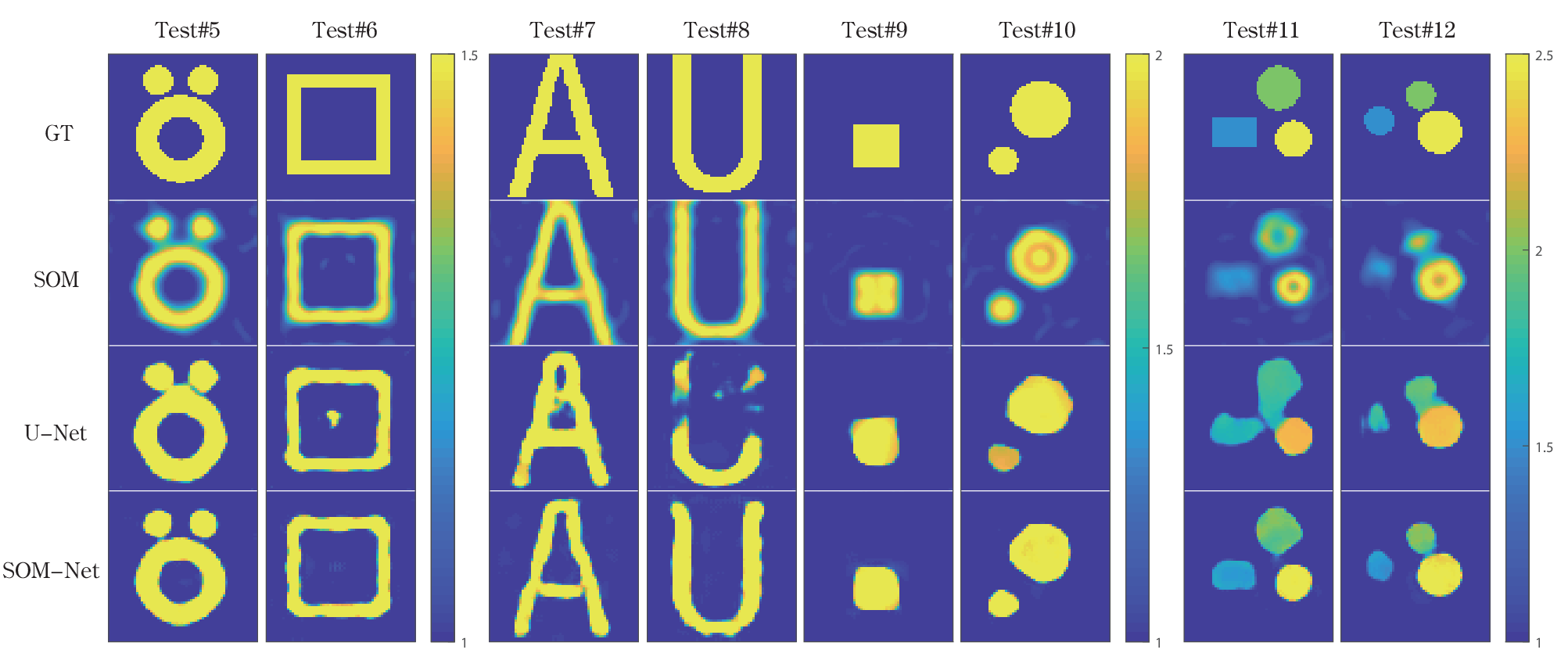}\\
  \caption{Reconstruction results of Test\#5 to Test\#12 for scatterers with complex profiles. }
  \label{compro}
\end{figure*}

\begin{table*}[ht]
\centering
\setlength{\abovecaptionskip}{-1pt}
\caption{Error metrics of reconstruction results for Test\#5-Test\#12.}
\setlength{\tabcolsep}{4pt}
\begin{tabular}{|c|cc|cc|cc|cc|cc|cc|cc|cc|}
\hline
\multirow{2}{*}{Method} & \multicolumn{2}{c|}{Test\#5}                       & \multicolumn{2}{c|}{Test\#6}                       & \multicolumn{2}{c|}{Test\#7}                       & \multicolumn{2}{c|}{Test\#8}                       & \multicolumn{2}{c|}{Test\#9}                       & \multicolumn{2}{c|}{Test\#10}                      & \multicolumn{2}{c|}{Test\#11}                      & \multicolumn{2}{c|}{Test\#12}                      \\ \cline{2-17}
                        & \multicolumn{1}{c|}{SSIM}          & RMSE          & \multicolumn{1}{c|}{SSIM}          & RMSE          & \multicolumn{1}{c|}{SSIM}          & RMSE          & \multicolumn{1}{c|}{SSIM}          & RMSE          & \multicolumn{1}{c|}{SSIM}          & RMSE          & \multicolumn{1}{c|}{SSIM}          & RMSE          & \multicolumn{1}{c|}{SSIM}          & RMSE          & \multicolumn{1}{c|}{SSIM}          & RMSE          \\ \hline
SOM                     & \multicolumn{1}{c|}{0.77}          & 0.07          & \multicolumn{1}{c|}{0.84}          & 0.07          & \multicolumn{1}{c|}{0.80}          & 0.13          & \multicolumn{1}{c|}{0.78}          & 0.13          & \multicolumn{1}{c|}{0.91}          & 0.07          & \multicolumn{1}{c|}{0.83}          & 0.10          & \multicolumn{1}{c|}{0.82}          & 0.11          & \multicolumn{1}{c|}{0.85}          & 0.12          \\ \hline
U-Net                   & \multicolumn{1}{c|}{0.84}          & 0.09          & \multicolumn{1}{c|}{0.85}          & 0.07          & \multicolumn{1}{c|}{0.76}          & 0.20          & \multicolumn{1}{c|}{0.69}          & 0.17          & \multicolumn{1}{c|}{0.94}          & 0.08          & \multicolumn{1}{c|}{0.92}          & 0.10          & \multicolumn{1}{c|}{0.84}          & 0.17          & \multicolumn{1}{c|}{0.90}          & 0.12          \\ \hline
SOM-Net                  & \multicolumn{1}{c|}{\textbf{0.93}} & \textbf{0.05} & \multicolumn{1}{c|}{\textbf{0.91}} & \textbf{0.05} & \multicolumn{1}{c|}{\textbf{0.83}} & \textbf{0.12} & \multicolumn{1}{c|}{\textbf{0.86}} & \textbf{0.10} & \multicolumn{1}{c|}{\textbf{0.97}} & \textbf{0.04} & \multicolumn{1}{c|}{\textbf{0.94}} & \textbf{0.08} & \multicolumn{1}{c|}{\textbf{0.92}} & \textbf{0.09} & \multicolumn{1}{c|}{\textbf{0.95}} & \textbf{0.07} \\ \hline
\end{tabular}
\label{tab2}
\end{table*}

\begin{table*}[ht]
\setlength{\abovecaptionskip}{-1pt}
\caption{Error metrics of reconstruction results for Test\#13-Test\#16.}
\centering
\setlength{\tabcolsep}{3mm}
\begin{tabular}{|c|cc|cc|cc|cc|}
\hline
\multirow{2}{*}{Method} & \multicolumn{2}{c|}{Test\#13}                      & \multicolumn{2}{c|}{Test\#14}                      & \multicolumn{2}{c|}{Test\#15}                      & \multicolumn{2}{c|}{Test\#16}                      \\ \cline{2-9}
                        & \multicolumn{1}{c|}{SSIM}          & RMSE          & \multicolumn{1}{c|}{SSIM}          & RMSE          & \multicolumn{1}{c|}{SSIM}          & RMSE          & \multicolumn{1}{c|}{SSIM}          & RMSE          \\ \hline
SOM                     & \multicolumn{1}{c|}{0.69}          & 0.15          & \multicolumn{1}{c|}{0.68}          & 0.15          & \multicolumn{1}{c|}{0.67}          & 0.16          & \multicolumn{1}{c|}{0.66}          & 0.16          \\ \hline
U-Net                   & \multicolumn{1}{c|}{0.80}          & 0.16          & \multicolumn{1}{c|}{0.77}          & 0.17          & \multicolumn{1}{c|}{0.76}          & 0.18          & \multicolumn{1}{c|}{0.77}          & 0.19          \\ \hline
SOM-Net                  & \multicolumn{1}{c|}{\textbf{0.90}} & \textbf{0.11} & \multicolumn{1}{c|}{\textbf{0.88}} & \textbf{0.11} & \multicolumn{1}{c|}{\textbf{0.88}} & \textbf{0.12} & \multicolumn{1}{c|}{\textbf{0.86}} & \textbf{0.13} \\ \hline
\end{tabular}
\label{tab3}
\end{table*}

\section{Numerical and experimental results}
In this section, we conduct  verifications of  SOM-Net with  both synthetic  and experimental examples. The performance of the SOM-Net is compared with the classical SOM method and U-Net \cite{wei2018deep}. The comparison with SOM indicates the benefit of the improvement on reconstruction quality and speed of SOM-Net, while the comparison with U-Net denotes the superior performance of SOM-Net over existing physics-inspired deep learning methods.

\subsection{Configuration of the Scattering System}
The configuration of synthetic  ISPs  is defined as follows. 16 line sources and 32 line receivers are equally placed on a circle with a radius of 3 m centered at the origin. The operating frequency is 400 MHz. For each transmitting antenna, all receivers measure the scattered field. {By default, each scatterer is assumed to be lossless if not mentioned. And the inversion of lossy scatterers  will be  described in the discussions.}  The domain $D$ has the size of 2 m $\times$ 2 m in a free space background. It is discretized into $100\times100$ grids to simulate the measured scattered field by MoM, while the discretization is changed to $64\times64$ grids in inversion to avoid the inverse crime. Suppose that the measured scattered fields of all receivers from all transmitters are recorded in a matrix $\bar{\bar{{E}}}^{s}$ with the size of $N_{r} \times N_{i}$.   $\bar{\bar{{E}}}^{s}$ is noise-free in the training stage, while Gaussian noise is added to $\bar{\bar{{E}}}^{s}$ in the testing stage to validate the robustness of SOM-Net against noise.  The noise level is defined as ($ \left \|\bar{\bar{n}}  \right \|_{F} / \left \|\bar{\bar{E}}^{s}\right \|_{F} $), where $\left \| \cdot \right \|_{F}$ represents Frobenius norm of a matrix.

\subsection{Training Details}
The training details of the SOM-Net are as follows.  The MNIST data set \cite{lecun1998gradient} is used to generate the  training data. Considering the complexity of the ISP, we randomly choose 5000 handwritten digits  in the MNIST database for  training  and another 2500 digits for validation. In order to increase the richness of data and enhance the generalization ability of the model,  each digit is randomly rotated with an angle of $[-170^{\circ},  170^{\circ}]$ degrees. Besides,  a circle with a random radius between 0.1 m to 0.5 m is also added in $D$. The relative permittivity of all training scatterers is set between 1.5 and 2.5. Here ${L}= 15$ is used in \eqref{Eq7} to generate deterministic induced current following the criterion of \cite{wei2018deep}.

The hyperparameters for training the SOM-Net are set as follows. {Both the $\lambda _{1}$ and $\lambda _{2}$ are experimentally set to 2.0.} The Adam optimizer is taken to optimize the SOM-Net with $\beta _{1}=0.9$ and $\beta _{2}=0.999$. The batch size is set to 16, and the training epochs are set to 40. The first 20 epochs are running with the same learning rate of 0.0002, while the learning rate decays linearly to 0 from the 21st to the 40th epochs. The code is implemented using PyTorch on a server with Intel(R) Core(TM) i9-10900X CPU @3.70GHz and GeForce RTX 3090 GPU. {The training process of the SOM-Net took about 7.26 hours under the above configurations.}

\begin{figure*}[ht]
  \centering
  \setlength{\abovecaptionskip}{-1pt}
  \includegraphics[width=0.7\textwidth]{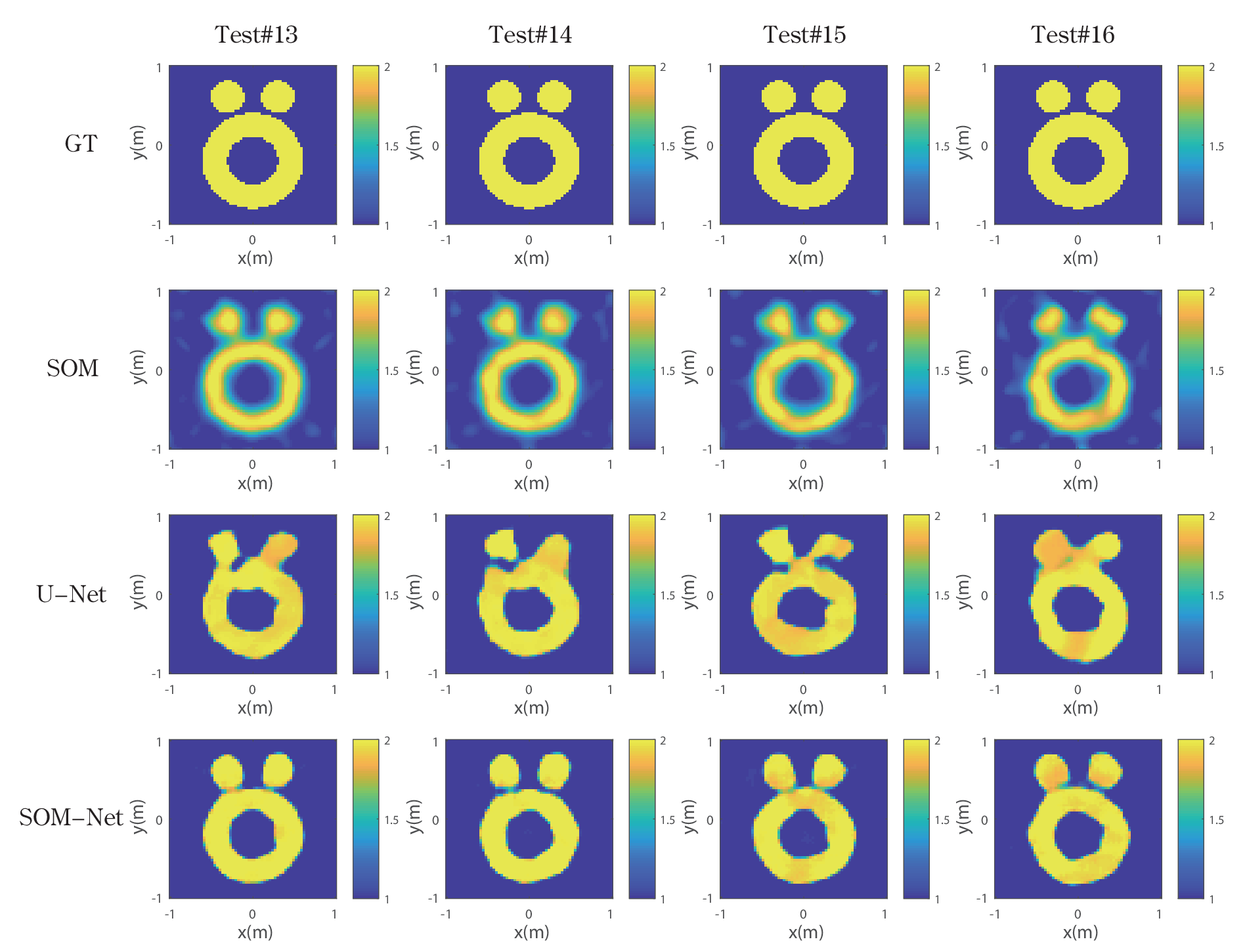}\\
  \caption{Reconstruction results of the ``Austria" profile  with the relative permittivity of 2.0 under different noise levels. From Test\#13 to Test\#16,  10\%, 20\%, 25\% and 30\% white Gaussian noise are added to the scattered field, respectively. }
  \label{austria2}
\end{figure*}

\subsection{Within-Database Test: Synthetic Data}
In the first example, we randomly select another 1500 images from MNIST database as the testing set. All the scattered fields of testing samples are added with $10\% $ White Gaussian  noise. The relative permittivity of all testing scatterers is between 1.5 and 2.5. The reconstruction results of four randomly selected examples, i.e., Test\#1 to Test\#4, are shown in Fig. \ref{digits4}.  We take the structural similarity measure (SSIM) and the root-mean-square error (RMSE) of the relative permittivity of scatterers as the quality metrics \cite{song2021electromagnetic} to evaluate the performance of all methods. The results are summarized in Table \ref{tab1}. It can be seen that the SOM-Net achieves superior performance compared to other two methods.

Since the SOM-Net is trained by enforcing the consistency of the induced current, the scattered field, and the permittivity images, Fig. \ref{digits4J}  shows the deterministic induced current, the predicted induced current distribution, and the corresponding target one for Test\#1 to Test\#4. The induced currents of other transmitters are similar and are omitted here. Meanwhile, the predicted scattered {field}  and its reference with respect to all incidences are plotted in Fig. \ref{digits4ES}, where the predicted field is represented by the orange dots and the reference one is indicated by the blue line. It is observed that the physical constraints of SOM-Net on the induced current and the scattered field make all those variables consistent with their references. Therefore, the SOM-Net learns the governing  physical law and achieves much better reconstruction results compared to U-Net.

Considering the simulation time, the SOM-Net only takes about 0.53 seconds to reconstruct a single result with given inputs. Specifically,  it takes about 0.03 seconds to obtain the rough contrast by BP, and another 1.4 seconds to obtain the deterministic induced current $\bar{J}^{+}$. So, the whole simulation time of SOM-Net to reconstruct one example is about 1.96 seconds. In comparison, the SOM method needs about 27.5 seconds for 50 iterations of reconstruction. The simulation time has been significantly reduced by 92\% with the SOM-Net.

\begin{figure}[!t]
  \setlength{\abovecaptionskip}{-12pt}
  \includegraphics[width=0.45\textwidth]{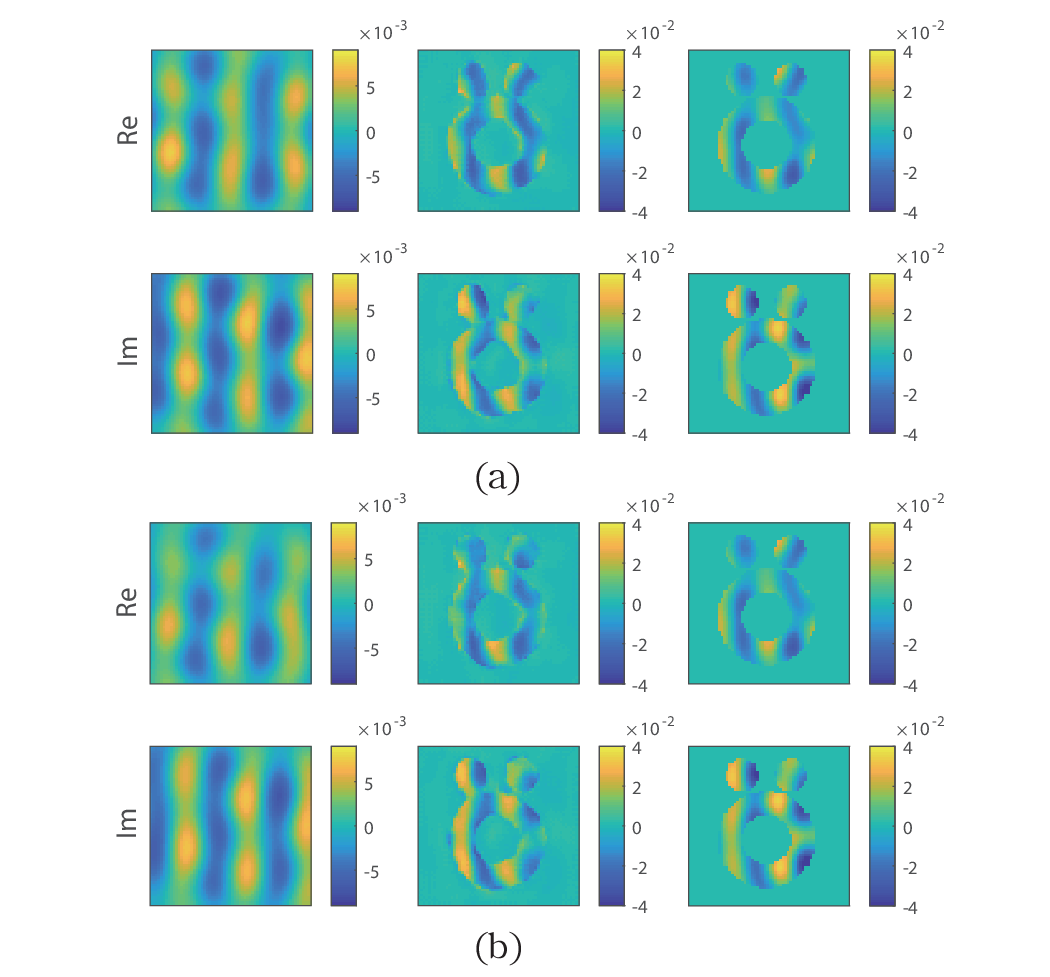}\\
  \caption{The fitting of  induced currents by the first transmitter for Test\#15 and Test\#16: (a) Test\#15, (b) Test\#16. For each test,  the left, middle, and right columns are the deterministic induced current, the output induced current of SOM-Net, and the target induced current, respectively. And the first row is the real part (Re), and the second row is the imaginary part Im).}
  \label{austria2J}
\end{figure}

\begin{figure}[!t]
\setlength{\abovecaptionskip}{-12pt}
  \includegraphics[width=0.47\textwidth]{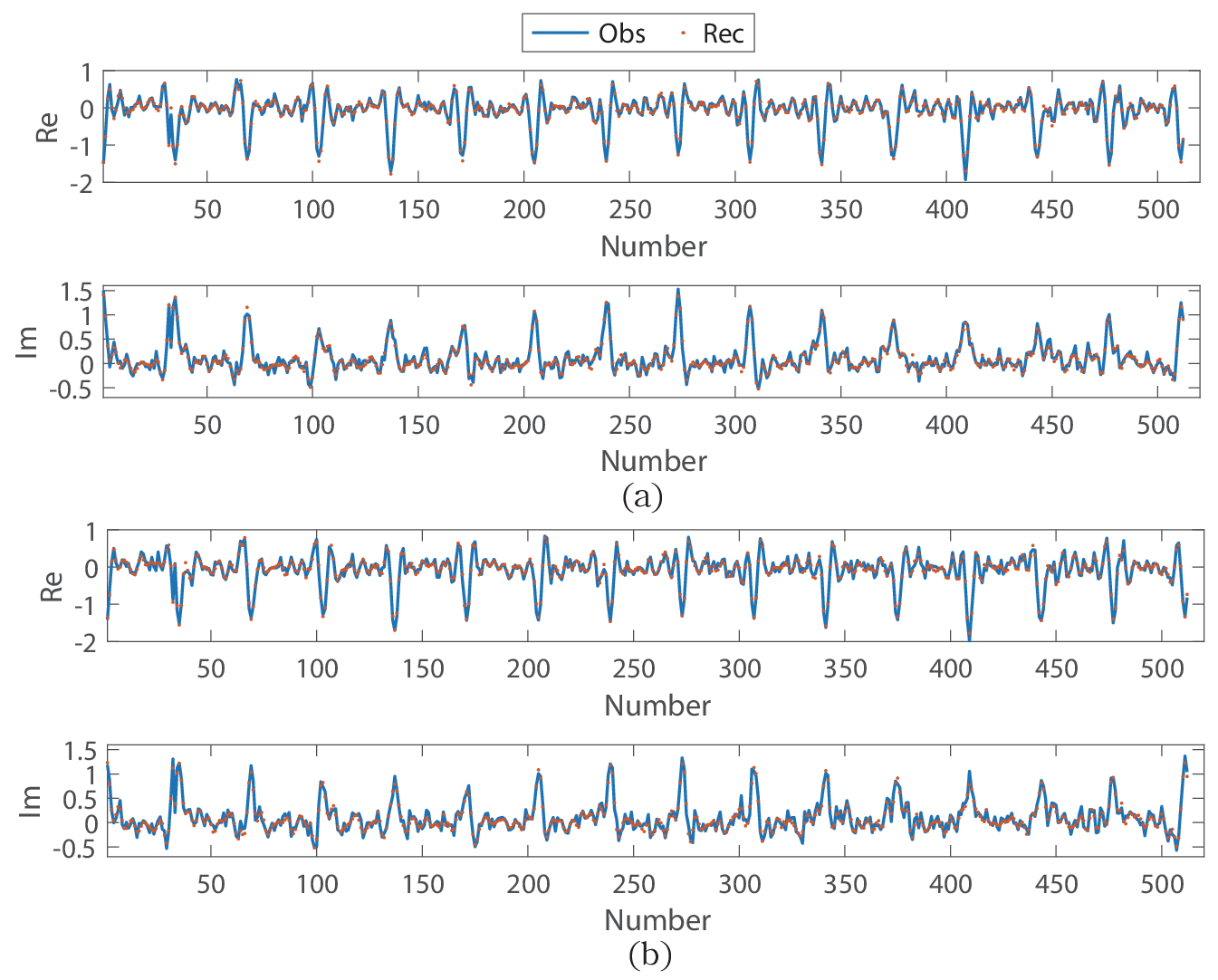}\\
  \caption{The fitting of all scattered fields for Test\#15 and Test\#16: (a) Test\#15, (b) Test\#16. For each test, the predicted scattered field by SOM-Net is represented by the orange dots, and the reference one is indicated by the blue line. }
  \label{austria2ES}
\end{figure}

\subsection{Cross-Database Test: Synthetic Data}
To verify the generalization ability of SOM-Net, we also test the SOM-Net with more challenging examples,  including  scatterers with complex shapes, high-level noise interference, and scatterers
with high relative permittivity. We still use the same testing model as the MNIST examples.

\begin{figure*}[!t]
  \centering
  \setlength{\abovecaptionskip}{-6pt}
  \includegraphics[width=0.7\textwidth]{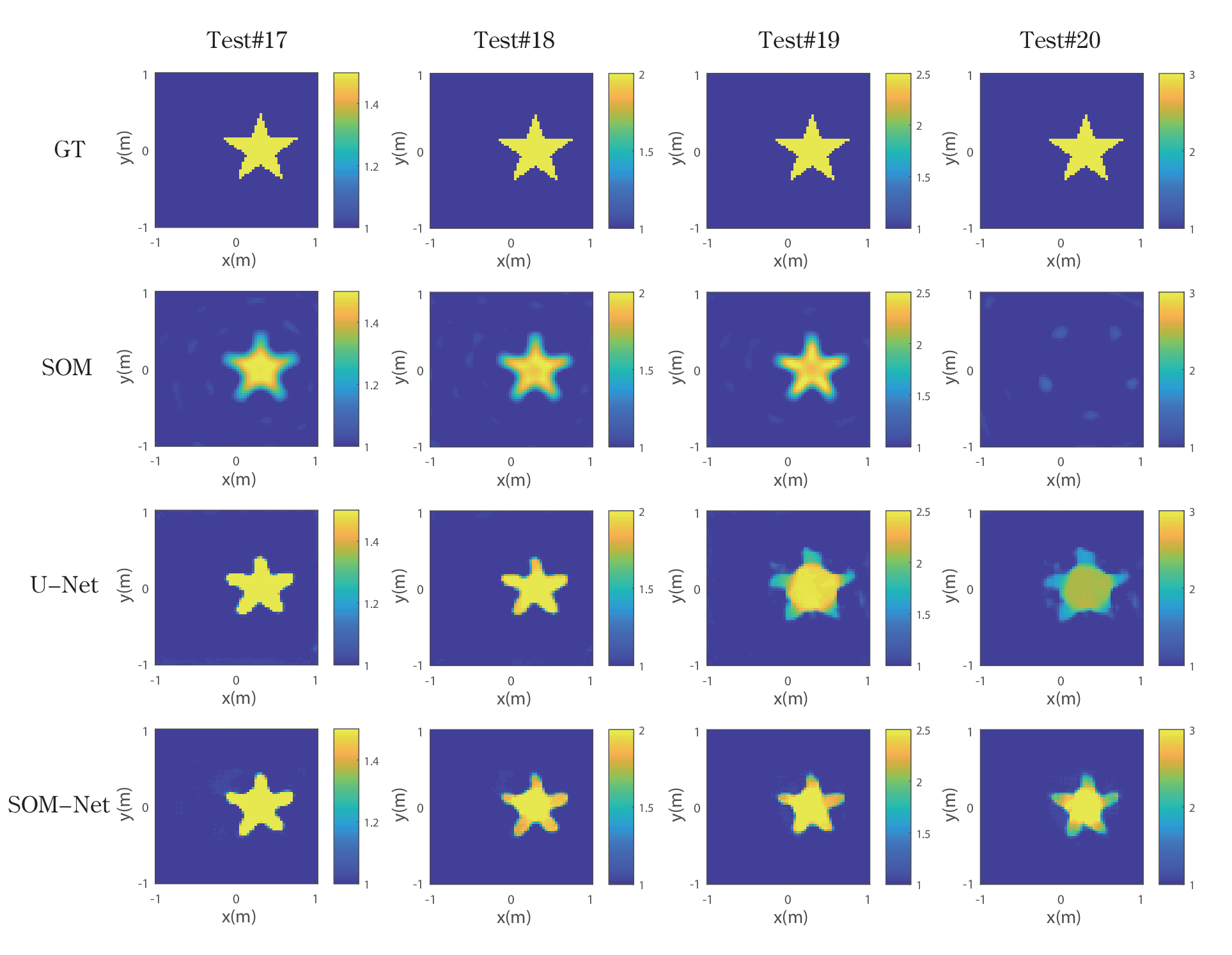}\\
  \caption{Reconstruction results of the star profile with the relative permittivity of 1.5, 2.0, 2.5, and 3.0, respectively. }
  \label{star}
\end{figure*}

\begin{table*}[]
\setlength{\abovecaptionskip}{-1pt}
\caption{Error metrics of reconstruction results for Test\#17-Test\#20.}
\centering
\setlength{\tabcolsep}{3mm}
\begin{tabular}{|c|cc|cc|cc|cc|}
\hline
\multirow{2}{*}{Method} & \multicolumn{2}{c|}{Test\#17}                      & \multicolumn{2}{c|}{Test\#18}                      & \multicolumn{2}{c|}{Test\#19}                      & \multicolumn{2}{c|}{Test\#20}                      \\ \cline{2-9}
                        & \multicolumn{1}{c|}{SSIM}          & RMSE          & \multicolumn{1}{c|}{SSIM}          & RMSE          & \multicolumn{1}{c|}{SSIM}          & RMSE          & \multicolumn{1}{c|}{SSIM}          & RMSE          \\ \hline
SOM                     & \multicolumn{1}{c|}{0.92}          & \textbf{0.05}          & \multicolumn{1}{c|}{0.91}          & 0.09          & \multicolumn{1}{c|}{0.92}          & \textbf{0.11}          & \multicolumn{1}{c|}{0.78}          & 0.18          \\ \hline
U-Net                   & \multicolumn{1}{c|}{0.93}          & 0.06          & \multicolumn{1}{c|}{\textbf{0.94}} & \textbf{0.08}          & \multicolumn{1}{c|}{0.83}          & 0.23          & \multicolumn{1}{c|}{0.83}          & 0.24          \\ \hline
SOM-Net                  & \multicolumn{1}{c|}{\textbf{0.94}} & \textbf{0.05} & \multicolumn{1}{c|}{\textbf{0.94}} & \textbf{0.08} & \multicolumn{1}{c|}{\textbf{0.93}} & \textbf{0.11} & \multicolumn{1}{c|}{\textbf{0.92}} & \textbf{0.16} \\ \hline
\end{tabular}
\label{tab4}
\end{table*}

\subsubsection{Complex Profiles}
We test Test\#5-Test\#12 where scatterers have complex shapes different from the MNIST training data. All the scattered field are contaminated with $10\% $ white Gaussian noise. The reconstruction results of  Test\#5-Test\#12 are shown in Fig. \ref{compro}, where the quality metrics are summarized in Table \ref{tab2}. The results shows that the SOM-Net outperforms the comparison methods.
Particularly, in Test\#11 and Test\#12, although there are three different scatterers and scatterers have diverse permittivity values, the SOM-Net still accurately reconstructs all examples. The results effectively verify the good generalization ability of the SOM-Net, which is attributed to the unrolling model design and the physical loss constraints for model training.

\subsubsection{High-level Noise Interference}
We also  test the noise robustness of the SOM-Net model under different noise levels. In this testing, we choose the ``Austria" profile as the scatterer with the $\varepsilon _{r} $ as 2.0. And $10\%, 20\%, 25\%, $ and $ 30\%$ white Gaussian noise are respectively added to the scattered field. The reconstruction results of Test\#13-Test\#16 with different noise levels are shown in Fig. \ref{austria2}. The corresponding error metrics are summarized in Table \ref{tab3}. It demonstrates that the SOM-Net is still robust against noise, which achieves better reconstruction results compared to the other methods. It is worth noting that compared with the ``Austria'' test in existing unrolling ISP methods \cite{liu2021physical} and \cite{guo2021physics}, the SOM-Net obtains better imaging results, especially showing clear gaps between two small rings and large rings. It can be considered that the high-quality restoration of induced current plays an important role in the reconstruction of permittivity images.

The fitting of induced current and the scattered field are shown in Fig. \ref{austria2J} and Fig. \ref{austria2ES}, respectively. The induced current is still from the first transmitter and the one for other transmitters are similar. Fig. \ref{austria2J} and Fig. \ref{austria2ES} show that the SOM-Net can accurately predict the induced current and the scattered field for ``Austria" profile with  25\% and 30\%  high-level noise. These results verify  the noise robustness of the SOM-Net.

\subsubsection{Scatterers with Different Permittivities}
To further check the generalization ability of the SOM-Net, we also test the reconstruction performance of a star profile with varying permittivities. The results of Test\#17-Test\#20 are shown in Fig. \ref{star}, where the relative permittivity ranges from 1.5 to 3.0. And $10\%$ white Gaussian noise is added to all testing cases. The corresponding error metrics are summarized in Table \ref{tab4}. It is worth noting that the SOM-Net has shown superior performance even for the case of the relative permittivity of 3.0, where the shape and the acute angles of the scatterer can still be recovered. {The  explicit constraints on physical variables and the unrolling design of SOM-Net ensure that the network learns the ISP mapping with good generalizations.}

Although the SOM-Net successfully retrieves the high-permittivity scatterer, it is still very challenging to reconstruct scatterers with a quite high permittivity. One potential way to overcome this issue is to unroll the existing model that can handle the ISP with high permittivity. For example,  Zhong et al. \cite{zhong2019multiresolution}  has demonstrated that the contraction integral equation for inversion (CIE-I) is very effective to solve highly nonlinear ISP with high contrasts.  Sanghvi et al. \cite{sanghvi2019embedding} introduced a deep learning method to provide good initials  for the two-fold SOM (TSOM) to reconstruct scatterers with high permittivity. Dubey et al. \cite{dubey2022phaseless} introduces a new  Rytov approximation for solving high-permittivity ISP with phaseless data. It is meaningful to unroll such iterative methods  into neural networks to  retrieve   high-permittivity scatterers.

\begin{figure}[!htb]
  \centering
  \includegraphics[width=0.25\textwidth]{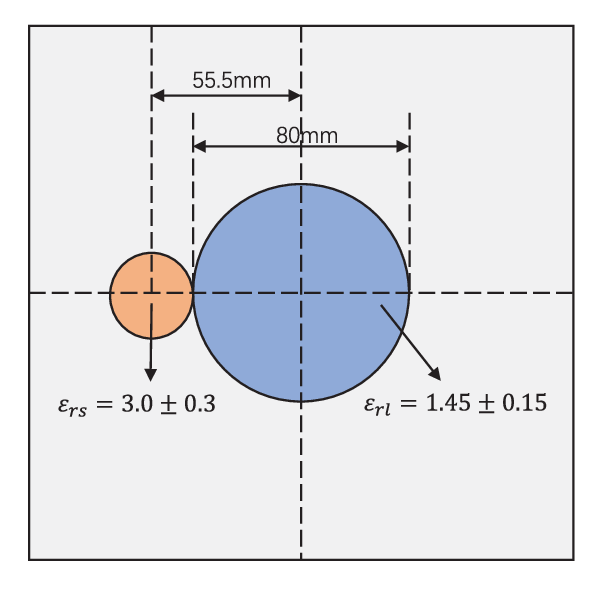}\\
  \caption{The ``FoamDielExt" profile from Fresnel experimental data.}
  \label{Ext}
\end{figure}

\begin{figure}[!h]
  \includegraphics[width=0.5\textwidth]{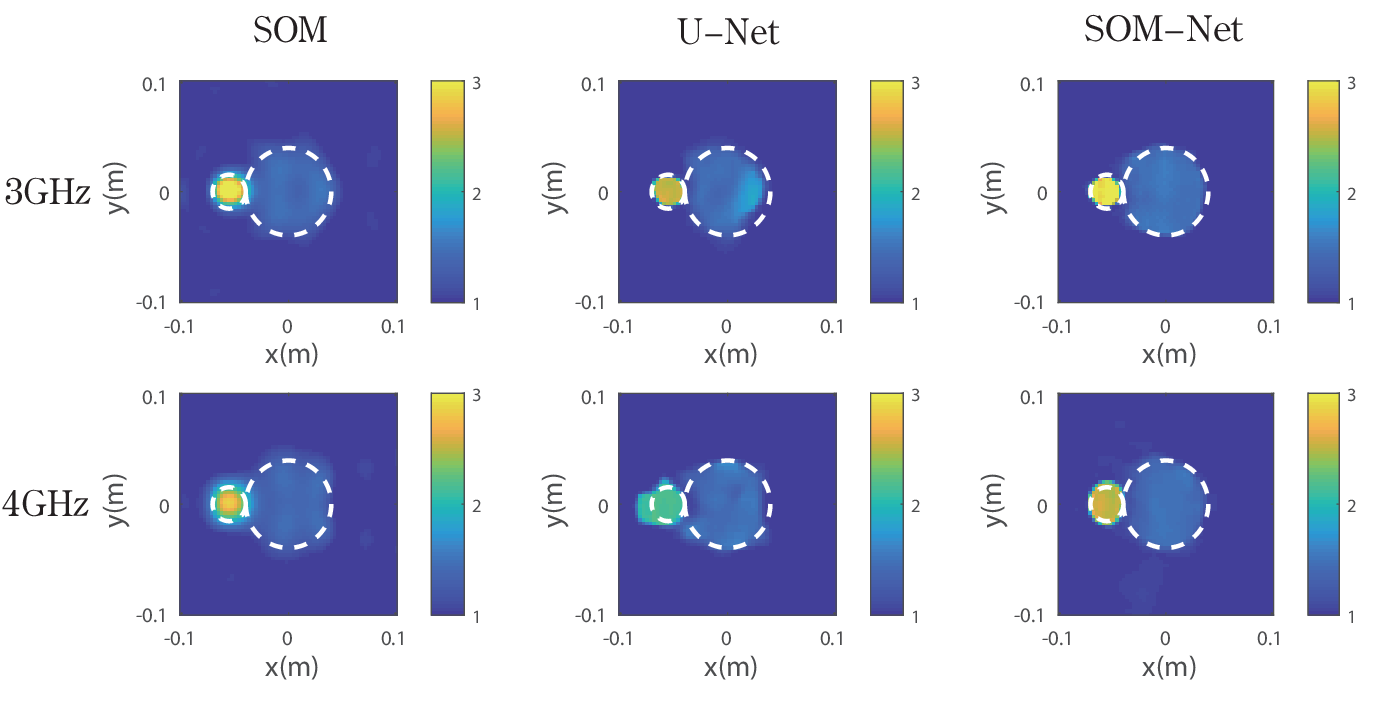}\\
  \setlength{\abovecaptionskip}{-10pt}
  \caption{Reconstruction results of the ``FoamDielExt" profile at 3GHz and 4GHz, respectively.}
  \label{Ext3+4}
\end{figure}

\begin{table}[!h]
\setlength{\abovecaptionskip}{-1pt}
\caption{Error metrics of reconstruction results in Fig.\ref{Ext3+4} for experimental data at 3GHz and 4GHz, respectively.}
\centering
\setlength{\tabcolsep}{3mm}
\begin{tabular}{|c|cc|cc|}
\hline
f/GHz   & \multicolumn{2}{c|}{3.0}                            & \multicolumn{2}{c|}{4.0}                            \\ \hline
Method & \multicolumn{1}{c|}{SSIM}          & RMSE          & \multicolumn{1}{c|}{SSIM}          & RMSE          \\ \hline
SOM    & \multicolumn{1}{c|}{0.91}          & 0.09          & \multicolumn{1}{c|}{0.88}          & \textbf{0.12} \\ \hline
U-Net  & \multicolumn{1}{c|}{0.91}          & \textbf{0.07} & \multicolumn{1}{c|}{0.89}          & 0.15          \\ \hline
SOM-Net & \multicolumn{1}{c|}{\textbf{0.94}} & \textbf{0.07} & \multicolumn{1}{c|}{\textbf{0.92}} & 0.13          \\ \hline
\end{tabular}
\label{tab5}
\end{table}

\subsection{Cross-Database Test: Experimental Data}
We finally validate the proposed method with the experimental data provided by Fresnel Institute \cite{geffrin2005free}. As shown in Fig. \ref{Ext}, the ``FoamDielExt" profile  is considered in this section. It is composed of two cylinders, a foam cylinder with a diameter of 80 mm and  $\varepsilon _{r} =1.45\pm0.15$, and a plastic cylinder with a diameter of 31 mm and $\varepsilon _{r} =3.0\pm0.3$.
Different from the previous synthetic examples, there are 8 linear transmitters and 241 receivers to measure experimental data, which are located on a circle with a radius of $1.67m$ from the center of the target.

\begin{figure}[]
  \includegraphics[width=0.5\textwidth]{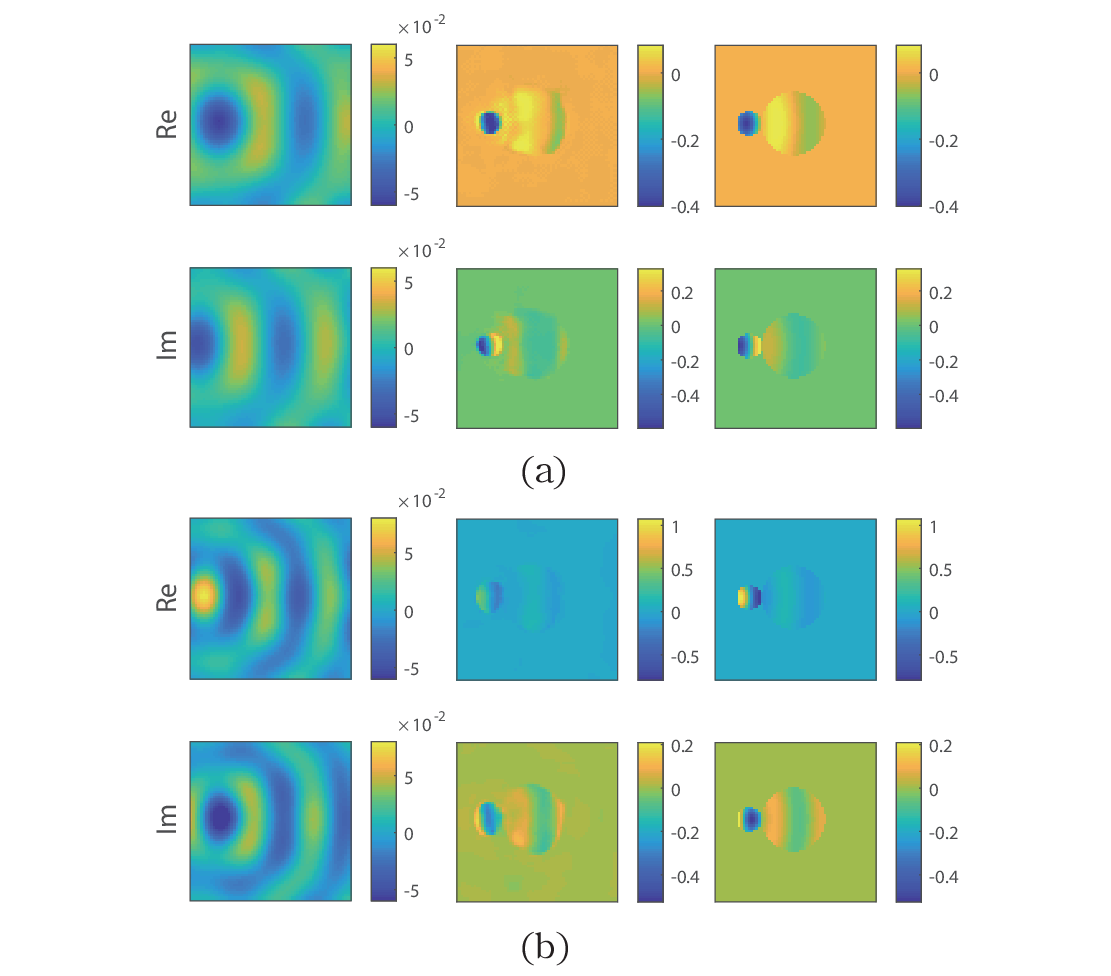}\\
  \setlength{\abovecaptionskip}{-10pt}
  \caption{The fitting of  induced currents by the first transmitter for ``FoamDielExt" profile at 3 GHz and 4 GHz, respectively: (a) 3 GHz, (b) 4 GHz. For each test,  the left, middle, and right columns are the deterministic induced current, the output induced current of SOM-Net, and the target induced current, respectively. And the first row is the real part (Re), and the second row is the imaginary part (Im). }
  \label{Ext3+4GJ}
\end{figure}

\begin{figure}[]
  \includegraphics[width=0.5\textwidth]{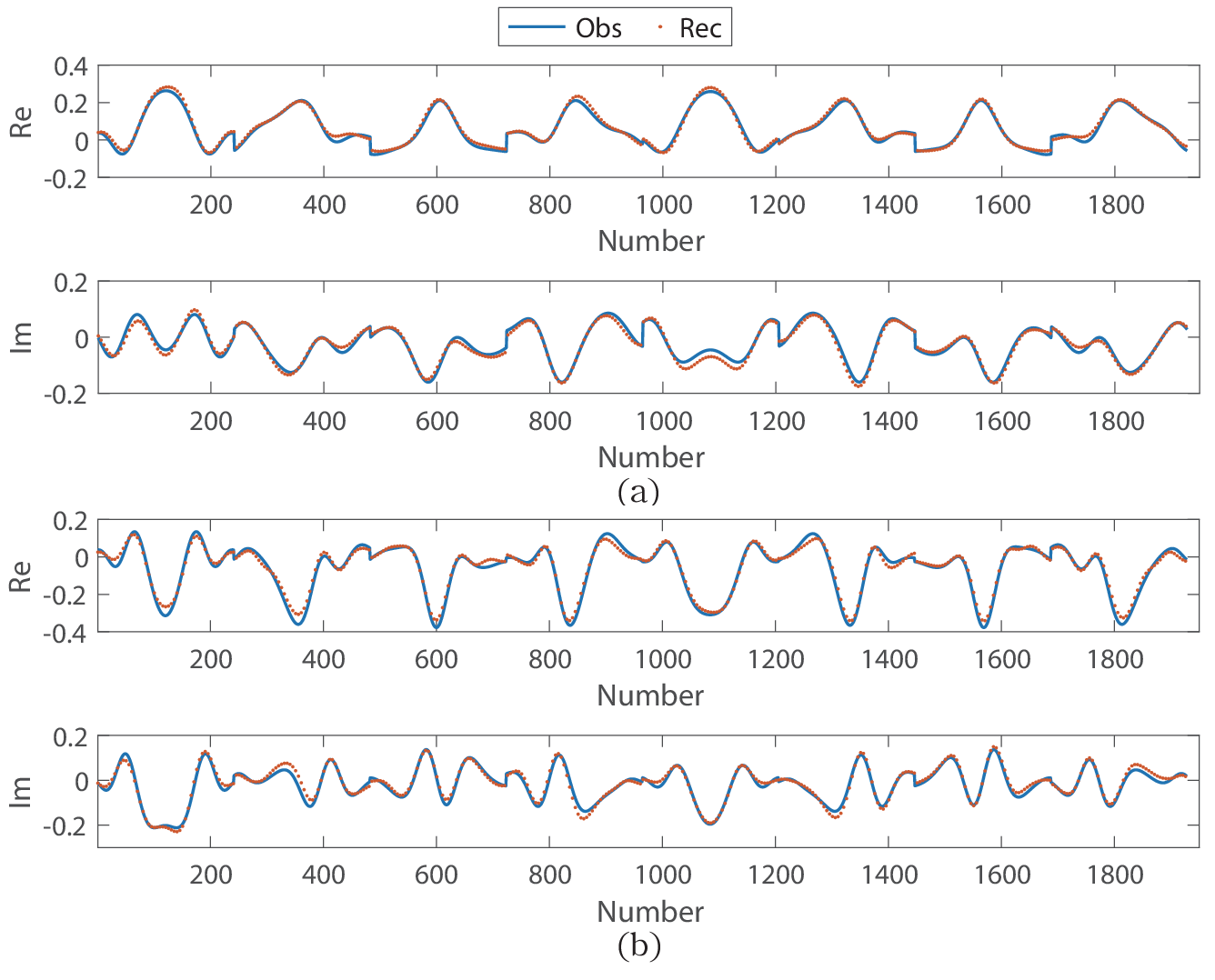}\\
  \setlength{\abovecaptionskip}{-12pt}
  \caption{The fitting of all scattered fields for the ``FoamDielExt" profile  at 3GHz and 4GHz, respectively: (a) 3 GHz, (b) 4 GHz. For each test,  the predicted scattered field by SOM-Net is represented by the orange dots every 5 points, and the reference one is indicated by the blue line.}
  \label{Ext3+4GES}
\end{figure}

The operating frequency is 3 GHz and 4 GHz, respectively, instead of the 400 MHz used in the synthetic examples. Accordingly, the size of DOI is also changed from 2 m $\times$ 2 m to 0.2 m $\times$ 0.2 m. In this experimental example, we use the same MNIST profiles with a random circle to generate the training data set, and the relative permittivity range of all scatterers is between 1.1 and 3.0.is
In Fig. \ref{Ext3+4}, we present the reconstruction profiles of SOM-Net, SOM, and U-Net, respectively.  The corresponding error metrics are summarized in Table \ref{tab5}. Specifically, the best SSIM is 0.94 at 3GHz, and 0.92 at 4GHz, respectively. The comparison results of the induced current and the scattered field are demonstrated in Fig. \ref{Ext3+4GJ} and Fig. \ref{Ext3+4GES}, respectively. The preferable reconstruction results once again verify the strong generalization ability of the SOM-Net.

\begin{figure}[!t]
  \centering
  \includegraphics[width=0.45\textwidth]{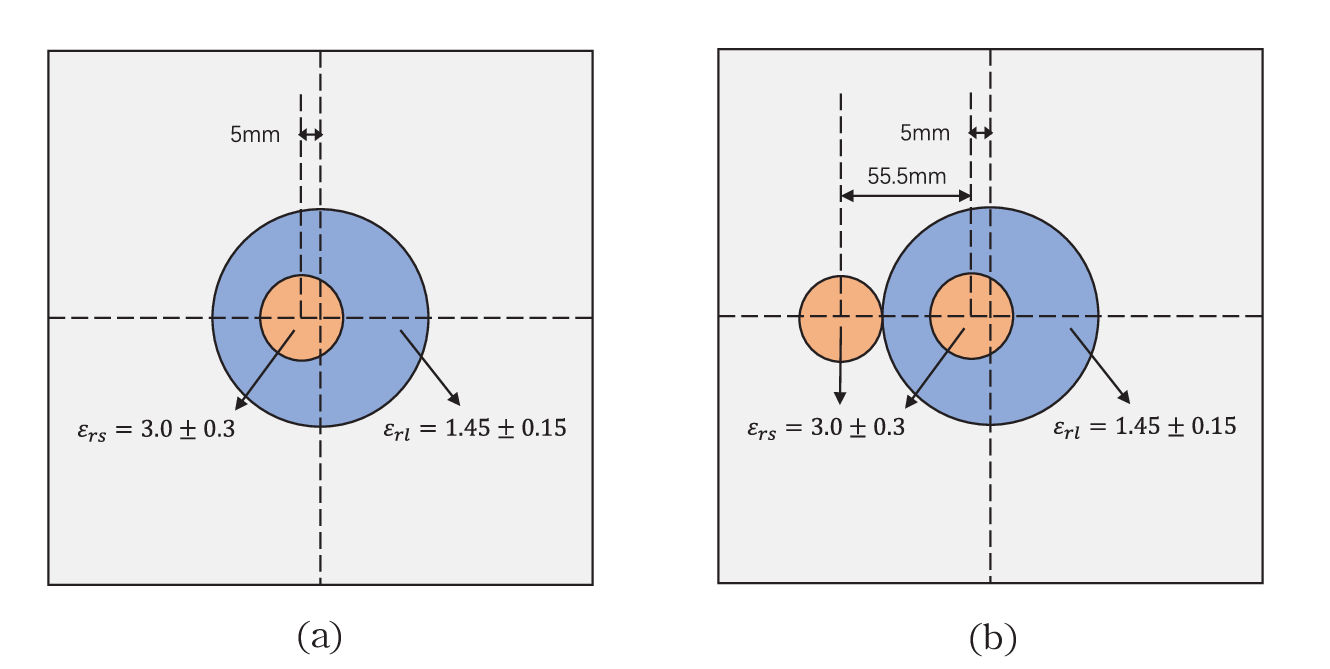}\\
  \setlength{\abovecaptionskip}{-1pt}
  \caption{ {The Fresnel experimental data: (a) the ``FoamDielInt" profile and (b) the ``FoamTwinDiel" profile.}}
  \label{FoamIntTwin}
\end{figure}

\begin{figure}[!htbp]
  \includegraphics[width=0.5\textwidth]{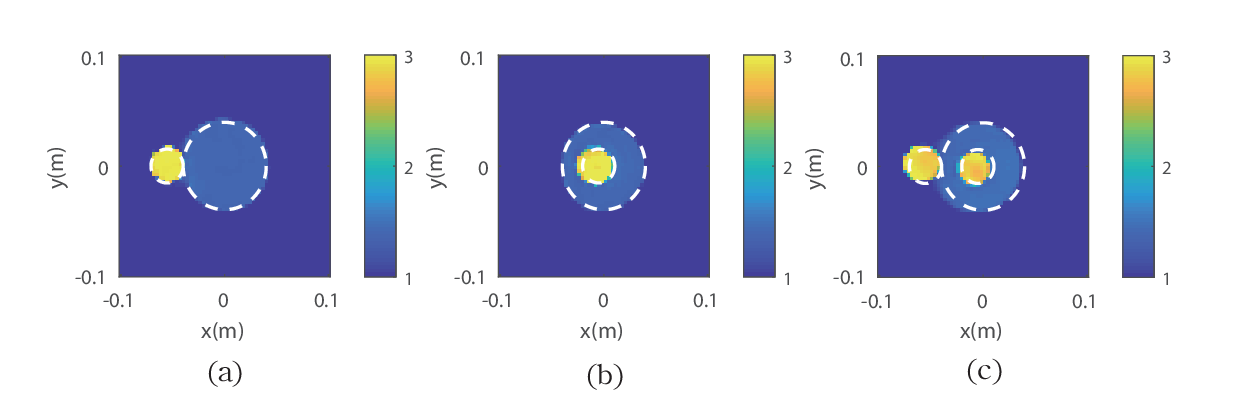}\\
  \setlength{\abovecaptionskip}{-16pt}
  \caption{{Reconstruction results of the experimental data: (a) the ``FoamDielExt" profile, (b) the ``FoamDielInt" profile, and (c) the ``FoamTwinDiel" profile at 4GHz, respectively.}}
  \label{Foamcom}
\end{figure}

\begin{table}[h]
\setlength{\abovecaptionskip}{-1pt}
\caption{{Error metrics of reconstruction results of SOM-Net in Fig. \ref{Foamcom} for experimental data at 4GHz.}}
\centering
\setlength{\tabcolsep}{3mm}
\begin{tabular}{|c|c|c|}
\hline
Test         & SSIM & RMSE \\ \hline
FoamDielExt  & 0.95 & 0.09 \\ \hline
FoamDielInt  & 0.95 & 0.08 \\ \hline
FoamTwinDiel & 0.90 & 0.16 \\ \hline
\end{tabular}
\label{FoamcomTab}
\end{table}

\section{Discussion}
In this section, we further discuss some aspects of the SOM-Net. It includes the comparison of SOM-Net with  existing unrolling methods \cite{liu2021physical,guo2021physics}, and the extension of SOM-Net for other cases like limited aperture and lossy scatterers.

\subsection{Comparison with Existing Unrolling Methods}
Here we take a further comparison of SOM-Net with  the existing unrolling work \cite{liu2021physical,guo2021physics} using the Fresnel's experimental data in Fig. \ref{Ext}  and  Fig. \ref{FoamIntTwin}.
In order to take a fair comparison, we employ the same configurations of generating training data as \cite{guo2021physics}, where each training sample consists of 2 or 3 random circles. More details can be seen in Table II in \cite{guo2021physics}. Specially, the predicted results of the SOM-Net are presented in Fig. \ref{Foamcom}, where both the reconstructed relative permittivity values and the shape are of high precision. Meanwhile, the corresponding SSIM and RMSE metrics  are summarized in Table \ref{FoamcomTab}. Considering the complexity to implement the methods in \cite{liu2021physical,guo2021physics}, we directly compare the SOM-Net results with those ones in the  two papers. In comparison, the reconstruction results of SOM-Net are visually superior to the  comparison algorithms. It should be noted that there is no quality metric given in \cite{liu2021physical}, and only the metric of data misfit ($rms_D$) on the scattered field is defined in \cite{guo2021physics}. Accordingly, the $rms_D$ of SOM-Net results for the ``FoamDielExt" and ``FoamDielInt" examples in Fig. \ref{Foamcom} (a) and (b) reaches 0.08 and 0.07, respectively, while that in \cite{guo2021physics} is 0.12 and 0.08, respectively. It shows that the predicted scattered field of the proposed method also matches better with the observed field.

\begin{figure*}[!t]
  \centering
  \includegraphics[width=1\textwidth]{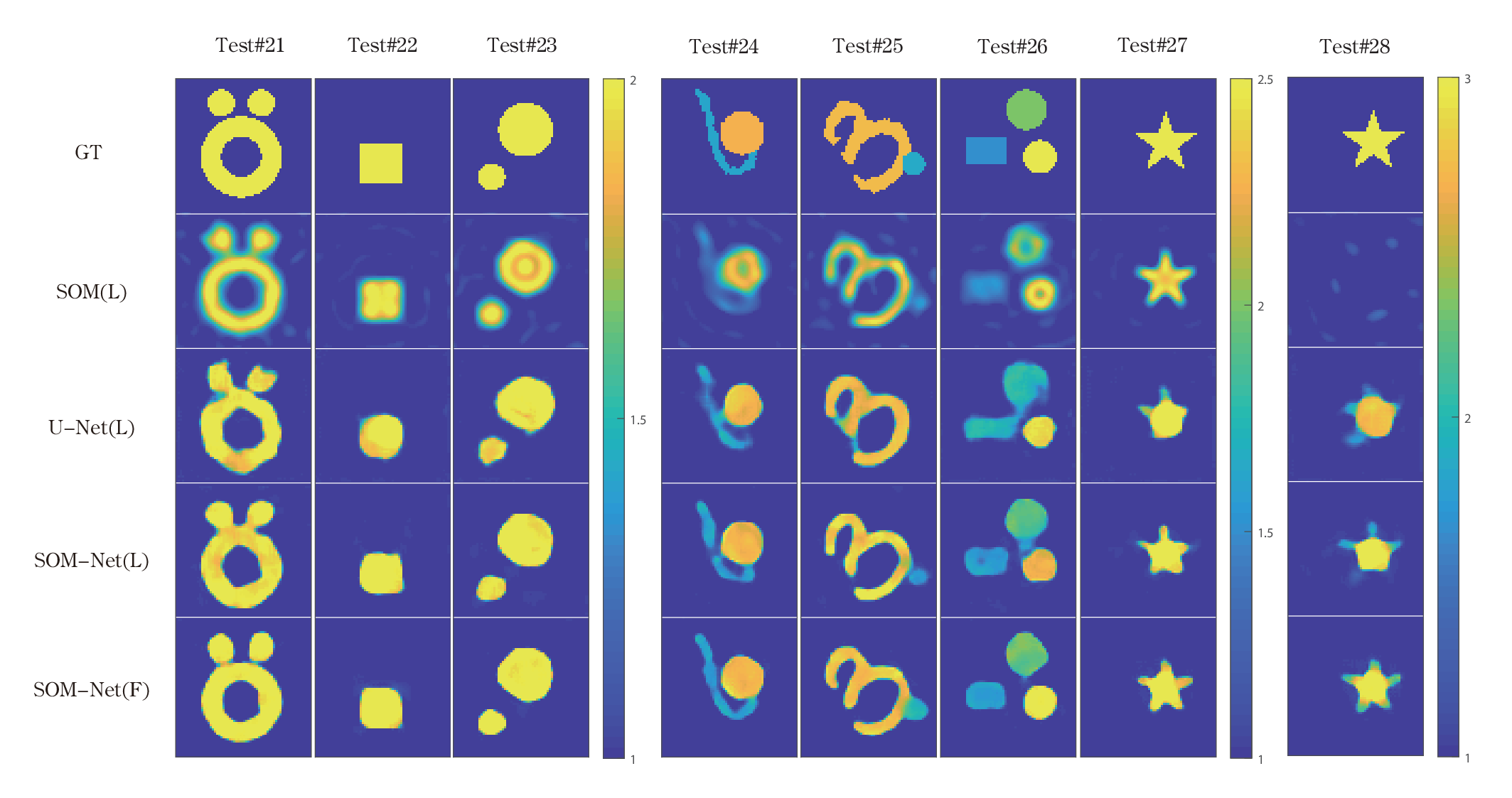}\\
  \setlength{\abovecaptionskip}{-4pt}
  \caption{{Reconstruction results of Test\#21 to Test\#28 for scatterers with 10\% white Gaussian noise.}}
  \label{Limiteder}
\end{figure*}

\begin{table*}[!tbp]
\setlength{\abovecaptionskip}{-1pt}
\caption{{Error metrics of reconstruction results for Test\#21-Test\#28 and 1500 MNIST tests.}}
\centering
\setlength{\tabcolsep}{2pt}
\begin{tabular}{|c|cc|cc|cc|cc|cc|cc|cc|cc|cc|}
\hline
\multirow{2}{*}{Method} & \multicolumn{2}{c|}{Test\#21}                      & \multicolumn{2}{c|}{Test\#22}                      & \multicolumn{2}{c|}{Tset\#23}                      & \multicolumn{2}{c|}{Test\#24}                      & \multicolumn{2}{c|}{Test\#25}                      & \multicolumn{2}{c|}{Test\#26}                      & \multicolumn{2}{c|}{Test\#27}                      & \multicolumn{2}{c|}{Test\#28}                      & \multicolumn{2}{c|}{1500 MNIST}                    \\ \cline{2-19}
                        & \multicolumn{1}{c|}{SSIM}          & RMSE          & \multicolumn{1}{c|}{SSIM}          & RMSE          & \multicolumn{1}{c|}{SSIM}          & RMSE          & \multicolumn{1}{c|}{SSIM}          & RMSE          & \multicolumn{1}{c|}{SSIM}          & RMSE          & \multicolumn{1}{c|}{SSIM}          & RMSE          & \multicolumn{1}{c|}{SSIM}          & RMSE          & \multicolumn{1}{c|}{SSIM}          & RMSE          & \multicolumn{1}{c|}{SSIM}          & RMSE          \\ \hline
SOM(L)                  & \multicolumn{1}{c|}{0.69}          & 0.15          & \multicolumn{1}{c|}{0.91}          & 0.07          & \multicolumn{1}{c|}{0.83}          & 0.10          & \multicolumn{1}{c|}{0.81}          & 0.12          & \multicolumn{1}{c|}{0.78}          & 0.16          & \multicolumn{1}{c|}{0.82}          & 0.11          & \multicolumn{1}{c|}{0.92}          & 0.11          & \multicolumn{1}{c|}{0.76}          & 0.18          & \multicolumn{1}{c|}{0.78}          & 0.14          \\ \hline
U-Net(L)                & \multicolumn{1}{c|}{0.79}          & 0.15          & \multicolumn{1}{c|}{0.93}          & 0.07          & \multicolumn{1}{c|}{0.91}          & 0.10          & \multicolumn{1}{c|}{0.88}          & 0.10          & \multicolumn{1}{c|}{0.84}          & 0.18          & \multicolumn{1}{c|}{0.82}          & 0.15          & \multicolumn{1}{c|}{0.89}          & 0.15          & \multicolumn{1}{c|}{0.85}          & 0.24          & \multicolumn{1}{c|}{0.86}          & 0.13          \\ \hline
SOM-Net(L)              & \multicolumn{1}{c|}{{0.86}} & {0.13} & \multicolumn{1}{c|}{{0.96}} & {\textbf{0.04}} & \multicolumn{1}{c|}{{0.93}} & {\textbf{0.08}} & \multicolumn{1}{c|}{{0.91}} & {0.09} & \multicolumn{1}{c|}{{0.87}} & {0.13} & \multicolumn{1}{c|}{{0.91}} & {0.15} & \multicolumn{1}{c|}{{0.92}} & {\textbf{0.11}} & \multicolumn{1}{c|}{{0.90}} & {\textbf{0.15}} & \multicolumn{1}{c|}{{0.88}} & {0.12} \\ \hline
SOM-Net(F)              & \multicolumn{1}{c|}{\textbf{0.90}}          & \textbf{0.11}          & \multicolumn{1}{c|}{\textbf{0.97}}          & \textbf{0.04}         & \multicolumn{1}{c|}{\textbf{0.94}}          & \textbf{0.08}          & \multicolumn{1}{c|}{\textbf{0.95}}          & \textbf{0.07}          & \multicolumn{1}{c|}{\textbf{0.91}}          & \textbf{0.11}         & \multicolumn{1}{c|}{\textbf{0.92}}          & \textbf{0.09}          & \multicolumn{1}{c|}{\textbf{0.93}}          & \textbf{0.11}          & \multicolumn{1}{c|}{\textbf{0.92}}          & 0.16          & \multicolumn{1}{c|}{\textbf{0.91}}          & \textbf{0.10}          \\ \hline
\end{tabular}
\label{limitedtab}
\end{table*}

\begin{figure*}[!t]
  \centering
  \includegraphics[width=1\textwidth]{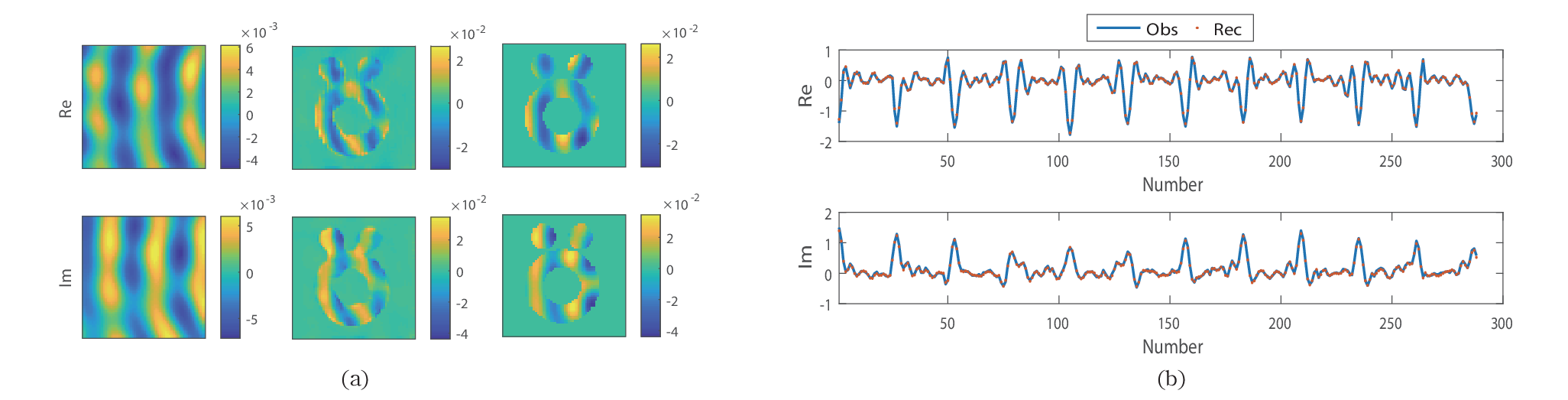}\\
  \setlength{\abovecaptionskip}{-4pt}
  \caption{{Fitting results of induced currents and scattered field in  Test\#21: (a) the fitting of induced currents by the first transmitter, where the left, middle, and right columns are the input, the output, and the target induced current, respectively. (b) the fitting of all scattered fields, where the predicted scattered field by SOM-Net is represented by the orange dots, and the reference one is indicated by the blue line.}}
  \label{LimitedausJEs}
\end{figure*}

\subsection{Limited Aperture}
Measurement within a limited aperture is a common ISP configuration in many real applications.  Here, we conduct the limited-aperture experiment using the same setup as the above full-aperture ones for synthetic examples, except that only measurements from transmitters and receivers within the observable aperture are used. Since available measurements are limited if the aperture is too small, we only consider  an experiment with a  $ 270^{\circ}$  aperture, where 12 line sources and 24 line receivers are available.

The reconstruction results with the $270^{\circ}$ aperture are demonstrated in Fig. \ref{Limiteder}, and the corresponding quality metrics are summarized in Table \ref{limitedtab}, where we use (L) and (F) to represent the results of limited and full apertures, respectively.
As observed, the proposed SOM-Net can still achieve better reconstruction performance for challenging cases than the SOM and U-Net. Meanwhile, the fitting results of the induced current by the first transmitter and the corresponding scattered fields for Test\#21 are shown in Fig. \ref{LimitedausJEs}, which indicates a good consistency with the references.

Meanwhile, it is also found that the reconstruction performance of SOM-Net degrades in the limited-aperture case compared to the full-aperture ones. {As shown in the comparison results of SOM-Net(L) and SOM-Net(F) in Fig. \ref{Limiteder} and Table \ref{limitedtab}, it is  observed that the SOM-Net with full-aperture data   achieves better reconstruction results compared to that of the SOM-Net(L). For example, there is a noticeable deterioration in the image quality  for Test\#21 in  the limited-aperture case. Meanwhile,  the SSIM decreases from 0.91 to 0.88, and the RMSE changes from 0.13 to 0.11. Similar quality  dropping   has also been observed in other examples. } This suggests that extra regularization  is required to improve the reconstruction of SOM-Net for dealing with such a challenging case, which needs further research but is beyond the scope of the current work.

\subsection{Lossy Profiles}
\begin{figure*}[!ht]
  \centering
  \includegraphics[width=0.95\textwidth]{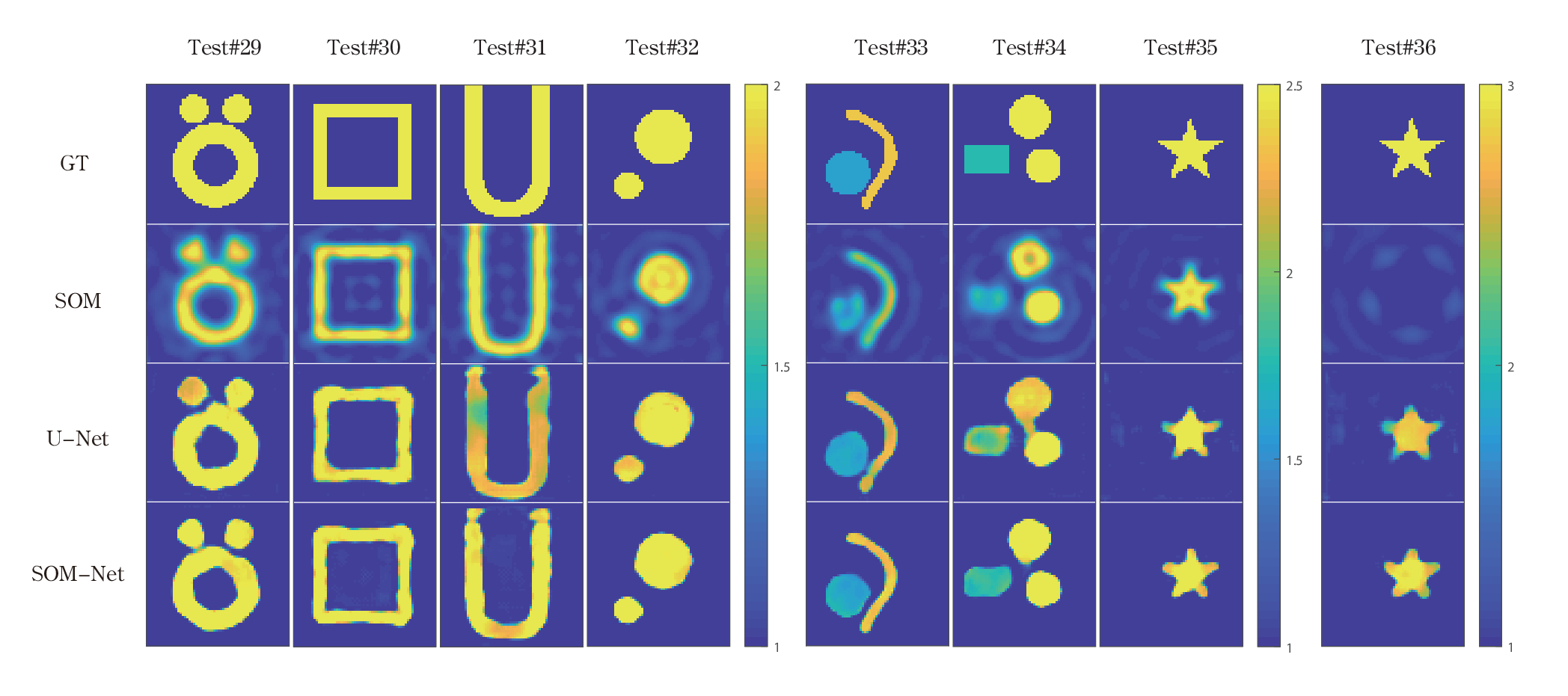}\\
  \setlength{\abovecaptionskip}{-8pt}
  \caption{{Reconstruction results of the real parts for lossy scatterers in Test\#29 to Test\#36 with 10\% white Gaussian noise. }}
  \label{Lossyreer}
\end{figure*}

\begin{figure*}[!t]
  \centering
  \includegraphics[width=0.95\textwidth]{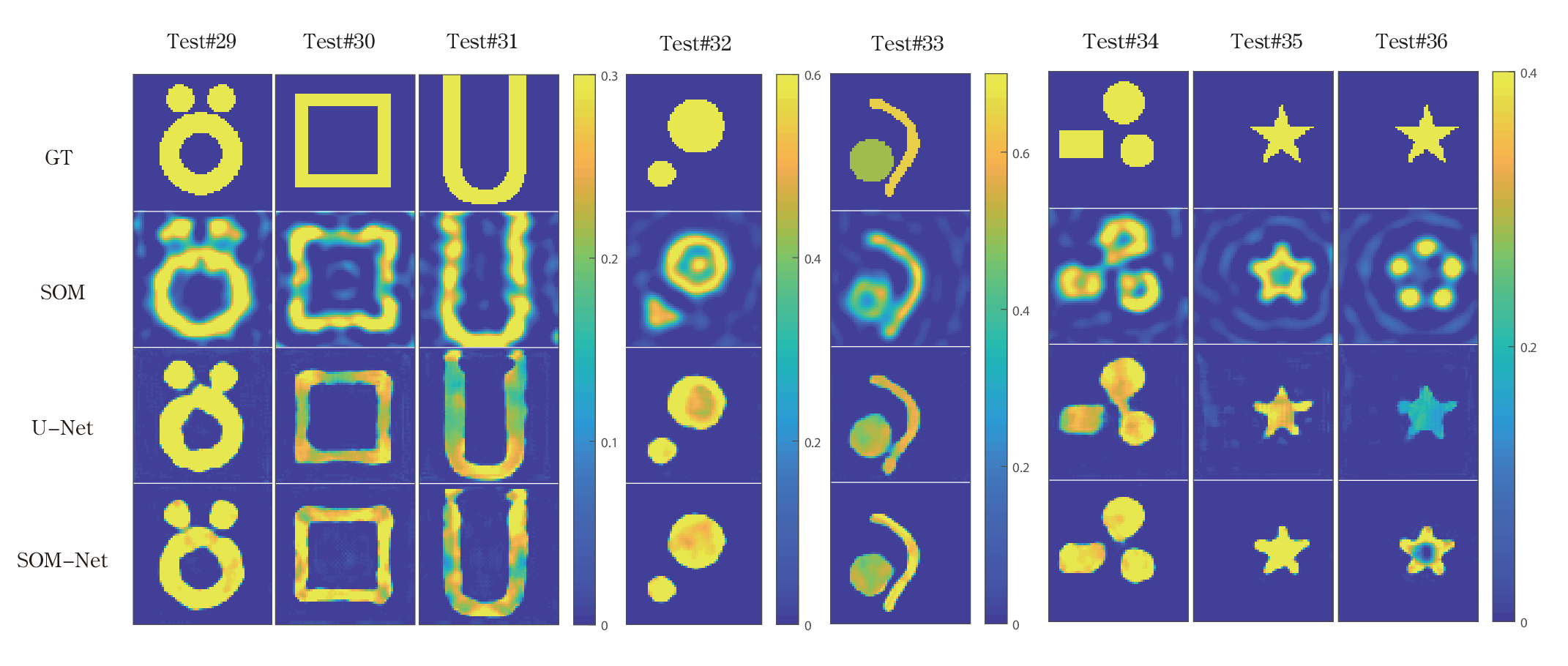}\\
  \setlength{\abovecaptionskip}{-5pt}
  \caption{{ Reconstruction results of the imaginary parts for lossy scatterers in Test\#29 to Test\#36 with 10\% white Gaussian noise.}}
  \label{Lossyimer}
\end{figure*}

Finally, we verify the performance of the  SOM-Net for the reconstruction of lossy profiles. Accordingly, we still conduct the lossy experiments using the same configurations as the synthetic lossless ones.  The only difference is that each scatterer here has an imaginary part randomly defined between 0 and 0.9. In addition, another channel representing the imaginary part of the complex relative permittivity image for lossy profiles also needs to be added to the SOM-Net. The reconstruction results of the real and imaginary parts for some typical samples are shown in Fig. \ref{Lossyreer} and Fig. \ref{Lossyimer}, respectively. The corresponding quality metrics are listed in Table \ref{lossytab}.

According to the reconstruction results, it can be seen that the average results of 1500 lossy samples in the MNIST data set verify the superior performance of SOM-Net over the compared ones. In particular, as observed in Fig. \ref{Lossyreer} and Fig. \ref{Lossyimer}, the proposed SOM-Net can still reconstruct challenging lossy profiles in high quality. In Fig. \ref{LossyausJEs}, the fitting results of the induced current and all scattered fields for the ``Austria" profile are drawn to further demonstrate the effective learning of physical knowledge. Note that in the presentation of concrete results, the first row is the real part (Re) and the second row is the imaginary part (Im) for the induced current and the scattered field.

It should be mentioned that, although we only describe the SOM-Net with a TM case, considering the similarity of the iterative SOM algorithms for the TM \cite{chen2009subspace} and transverse electric (TE) \cite{pan2009application} cases, it is also easy to extend the SOM-Net to the TE case. However, the computational complexity will increase quickly compared to the TM case.

\begin{table*}[htbp]
\setlength{\abovecaptionskip}{-1pt}
\caption{{Error metrics of reconstruction results for Test\#29-Test\#36 and 1500 MNIST tests.}}
\centering
\setlength{\tabcolsep}{4.5pt}
\begin{tabular}{|c|cc|cc|cc|cc|cc|cc|cc|cc|cc|}
\hline
Test    & \multicolumn{2}{c|}{Test\#29}                      & \multicolumn{2}{c|}{Test\#30}                      & \multicolumn{2}{c|}{Tset\#31}                      & \multicolumn{2}{c|}{Tset\#32}                      & \multicolumn{2}{c|}{Tset\#33}                      & \multicolumn{2}{c|}{Tset\#34}                      & \multicolumn{2}{c|}{Tset\#35}                      & \multicolumn{2}{c|}{Tset\#36}                      & \multicolumn{2}{c|}{1500 MNIST}                    \\ \hline
SSIM    & \multicolumn{1}{c|}{Re}            & Im            & \multicolumn{1}{c|}{Re}            & Im            & \multicolumn{1}{c|}{Re}            & Im            & \multicolumn{1}{c|}{Re}            & Im            & \multicolumn{1}{c|}{Re}            & Im            & \multicolumn{1}{c|}{Re}            & Im            & \multicolumn{1}{c|}{Re}            & Im            & \multicolumn{1}{c|}{Re}            & Im            & \multicolumn{1}{c|}{Re}            & Im            \\ \hline
SOM     & \multicolumn{1}{c|}{0.64}          & 0.13          & \multicolumn{1}{c|}{0.67}          & 0.21          & \multicolumn{1}{c|}{0.68}          & 0.29          & \multicolumn{1}{c|}{0.77}          & 0.19          & \multicolumn{1}{c|}{0.76}          & 0.23          & \multicolumn{1}{c|}{0.74}          & 0.23          & \multicolumn{1}{c|}{0.86}          & 0.21          & \multicolumn{1}{c|}{0.61}          & 0.10          & \multicolumn{1}{c|}{0.69}          & 0.12          \\ \hline
U-Net   & \multicolumn{1}{c|}{\textbf{0.87}}          & 0.75          & \multicolumn{1}{c|}{0.89}          & 0.81          & \multicolumn{1}{c|}{0.75}          & 0.67          & \multicolumn{1}{c|}{0.95}          & 0.90          & \multicolumn{1}{c|}{0.94}          & 0.89          & \multicolumn{1}{c|}{0.89}          & 0.82          & \multicolumn{1}{c|}{\textbf{0.92}}          & 0.79          & \multicolumn{1}{c|}{0.87}          & 0.79          & \multicolumn{1}{c|}{0.89}          & 0.59          \\ \hline
SOM-Net & \multicolumn{1}{c|}{\textbf{0.87}} & \textbf{0.84} & \multicolumn{1}{c|}{\textbf{0.92}} & \textbf{0.86} & \multicolumn{1}{c|}{\textbf{0.89}} & \textbf{0.80} & \multicolumn{1}{c|}{\textbf{0.96}} & \textbf{0.94} & \multicolumn{1}{c|}{\textbf{0.95}} & \textbf{0.92} & \multicolumn{1}{c|}{\textbf{0.94}} & \textbf{0.91} & \multicolumn{1}{c|}{{0.91}} & \textbf{0.88} & \multicolumn{1}{c|}{\textbf{0.91}} & \textbf{0.88} & \multicolumn{1}{c|}{\textbf{0.90}} & \textbf{0.70} \\ \hline
\end{tabular}
\label{lossytab}
\end{table*}

\begin{figure*}[!htbp]
  \includegraphics[width=1\textwidth]{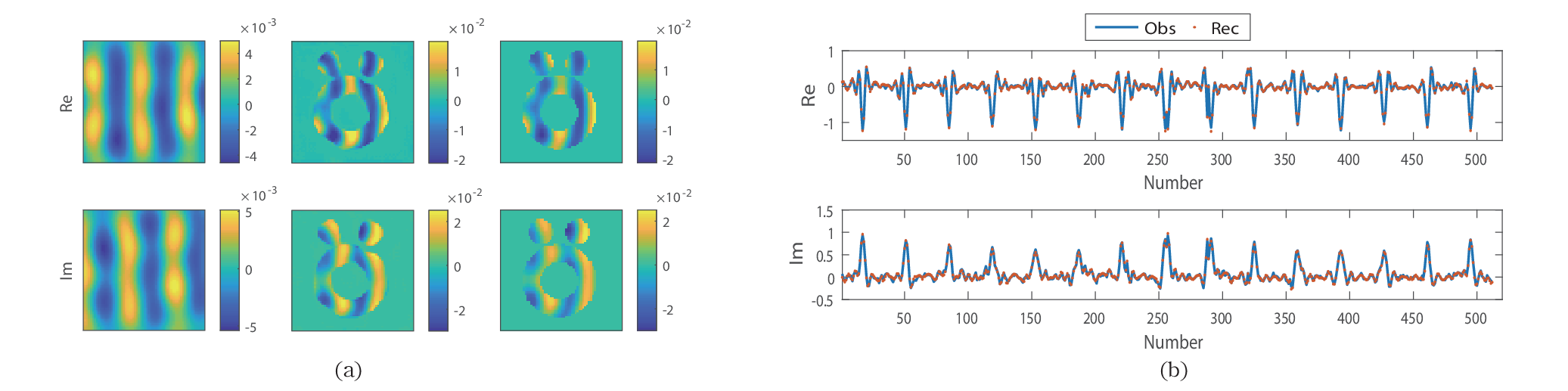}\\
  \setlength{\abovecaptionskip}{-14pt}
  \caption{{Fitting results of induced currents and scattered field in Test\#29: (a) the fitting of induced currents by the first transmitter, where the left, middle, and right columns are the input, the output, and the target induced current, respectively. (b) the fitting of all scattered fields, where the predicted scattered field by SOM-Net is represented by the orange dots, and the reference one is indicated by the blue line. }}
  \label{LossyausJEs}
\end{figure*}

\section{Conclusion}
In this work, we have proposed the SOM-Net for solving the full-wave ISPs by unrolling the iterative SOM. The SOM-Net has been  designed with several sub-networks to imitate the iterations of the SOM. The deterministic induced current and the rough permittivity image got by BP are used as the input, while the direct output of the SOM-Net is the full induced current. The induced current and the permittivity variables have been updated alternately in the cascade sub-networks of SOM-Net, which greatly reduces the nonlinearity of the ISP. Joint physical constraints of the induced current, the scattered field, and the permittivity image have been employed to guide the training of the SOM-Net model. The  generalization ability of the resulting SOM-Net model has been greatly enhanced, considering the inherent embedding of physical law on the network structure and  the  consistent matches of all physical variables of the governing Lippmann-Schwinger equations. All numerical tests have validated that the proposed SOM-Net has superior performance than SOM and U-Net. Overall, the SOM-Net has been verified to be a learning-based fast inversion method with strong generalization ability and good physical interpretability, which  provides a new idea for the application of deep unrolling technology to solve ISPs.

 Despite the good results achieved, the proposed SOM-Net still needs to be further improved and extended in many aspects, such as the inversion under a limited aperture, and the extension to the 3D case.

\bibliographystyle{IEEEtran}
\bibliography{reflib}

%\newpage
%\begin{IEEEbiography}[{\includegraphics[width=1in,height=1.25in,clip,keepaspectratio]{fig1}}]{Michael Shell}
%Use $\backslash${\tt{begin\{IEEEbiography\}}} and then for the 1st argument use $\backslash${\tt{includegraphics}} to declare and link the author photo.
%Use the author name as the 3rd argument followed by the biography text.
%\end{IEEEbiography}
%
%
%\vfill

\end{document}